\documentclass[11pt,twoside]{article}
\usepackage{amsmath}
\usepackage{amsfonts,amssymb,latexsym}

\begin{document}

\newcommand{\nl}{\nonumber\\}
\newcommand{\nnl}{\nl[6mm]}
\newcommand{\nle}{\nl[-2.5mm]\\[-2.5mm]}
\newcommand{\nlb}[1]{\nl[-2.0mm]\label{#1}\\[-2.0mm]}
\newcommand{\ab}{\allowbreak}

\renewcommand{\leq}{\leqslant}
\renewcommand{\geq}{\geqslant}

\renewcommand{\theequation}{\thesection.\arabic{equation}}
\let\ssection=\section
\renewcommand{\section}{\setcounter{equation}{0}\ssection}

\newcommand{\be}{\bes}
\newcommand{\ee}{\ees}
\newcommand{\bes}{\begin{eqnarray}}
\newcommand{\ees}{\end{eqnarray}}
\newcommand{\eens}{\nonumber\end{eqnarray}}
\newcommand{\barr}{\begin{array}}
\newcommand{\earr}{\end{array}}

\renewcommand{\/}{\over}
\renewcommand{\d}{\partial}
\newcommand{\Dslash}{\hbox{$D\kern-2.4mm/\,$}}
\newcommand{\dd}[1]{\ab\delta/\delta {#1}}
\newcommand{\ddt}{{d\/dt}}

\newcommand{\no}[1]{{\,:\kern-0.7mm #1\kern-1.2mm:\,}}

\newcommand{\mm}{{\mathbf m}}
\newcommand{\nn}{{\mathbf n}}
\newcommand{\cmm}{{,\mm}}
\newcommand{\cnn}{{,\nn}}

\newcommand{\xmu}{\xi^\mu}
\newcommand{\ynu}{\eta^\nu}
\newcommand{\qsmu}{q^*_\mu}
\newcommand{\qsnu}{q^*_\nu}
\newcommand{\psmu}{p_*^\mu}

\newcommand{\half}{{1\/2}}
\newcommand{\quart}{{1\/4}}
\newcommand{\tpi}{{1\/2\pi i}}
\newcommand{\bra}[1]{\big{\langle}#1\big{|}}
\newcommand{\ket}[1]{\big{|}#1\big{\rangle}}

\newcommand{\da}{\d_\alpha}
\newcommand{\db}{\d_\beta}
\newcommand{\dc}{\d_\gamma}
\newcommand{\za}{\zeta_a}
\newcommand{\zb}{\zeta_b}
\newcommand{\zc}{\zeta_c}
\newcommand{\Mab}{M^{\alpha\beta}}
\newcommand{\Kab}{K_{\alpha\beta}}

\newcommand{\fa}{\phi^\alpha}
\newcommand{\fb}{\phi^\beta}
\newcommand{\pa}{\pi_\alpha}
\newcommand{\pb}{\pi_\beta}
\newcommand{\Ea}{\EE_\alpha}
\newcommand{\Eb}{\EE_\beta}

\newcommand{\fs}{\phi^*}
\newcommand{\ps}{\pi_*}
\newcommand{\fsa}{\phi^*_\alpha}
\newcommand{\fsb}{\phi^*_\beta}
\newcommand{\fsc}{\phi^*_\gamma}
\newcommand{\psa}{\pi_*^\alpha}
\newcommand{\psb}{\pi_*^\beta}
\newcommand{\fsi}{\phi^*_i}

\newcommand{\fA}{\phi^A}
\newcommand{\fB}{\phi^B}
\newcommand{\pA}{\pi_A}
\newcommand{\wfA}{\bar\phi^A}
\newcommand{\wfB}{\bar\phi^B}

\newcommand{\As}{A_*}
\newcommand{\Es}{E^*}
\newcommand{\wA}{{\bar A}}
\newcommand{\wE}{{\bar E}}
\newcommand{\wAs}{{\bar A}{}_*}
\newcommand{\wEs}{{\bar E}{}^*}
\newcommand{\wc}{{\bar c}}
\newcommand{\wzeta}{{\bar\zeta}}
\newcommand{\wF}{{\bar F}}

\newcommand{\wf}{{\bar \phi}}
\renewcommand{\wp}{{\bar \pi}}
\newcommand{\wfs}{{\bar \phi}{}^*}
\newcommand{\wps}{{\bar \pi}_*}

\newcommand{\fm}{\phi_\cmm}
\newcommand{\dotfm}{\dot\phi_\cmm}
\newcommand{\fn}{\phi_\cnn}
\newcommand{\pim}{\pi^\cmm}
\newcommand{\pin}{\pi^\cnn}
\newcommand{\fsm}{\phi^*_\cmm}
\newcommand{\fsn}{\phi^*_\cnn}
\newcommand{\psm}{\pi_*^\cmm}
\newcommand{\psn}{\pi_*^\cnn}

\renewcommand{\fam}{\phi^\alpha_\cmm}
\newcommand{\dotfam}{\dot\phi^\alpha_\cmm}
\newcommand{\fbn}{\phi^\beta_\cnn}
\newcommand{\pam}{\pi_\alpha^\cmm}
\newcommand{\pbn}{\pi_\beta^\cnn}
\newcommand{\zam}{\zeta_{a\cmm}}
\newcommand{\zcm}{\zeta_{c\cmm}}
\newcommand{\cam}{c^a_\cmm}
\newcommand{\wfam}{{\bar \phi}{}^\alpha_\cmm}
\newcommand{\wfbn}{{\bar \phi}{}^\beta_\cnn}
\newcommand{\wpam}{{\bar \pi}{}_\alpha^\cmm}

\newcommand{\fAm}{\fA_\cmm}
\newcommand{\fAn}{\fA_\cnn}
\newcommand{\fBn}{\fB_\cnn}
\newcommand{\pAm}{\pA^\cmm}
\newcommand{\pAn}{\pA^\cnn}
\newcommand{\wfAm}{\bar\fA_\cmm}

\newcommand{\wfm}{{\bar \phi}_\cmm}
\newcommand{\wfn}{{\bar \phi}_\cnn}
\newcommand{\wpm}{{\bar \pi}^\cmm}
\newcommand{\wfsm}{\bar \phi^*_\cmm}
\newcommand{\wfsn}{\bar \phi^*_\cnn}
\newcommand{\wpsm}{\bar \pi_*^\cmm}
\newcommand{\wpsn}{\bar \pi_*^\cnn}

\newcommand{\wfa}{{\bar \phi}{}^\alpha}
\newcommand{\wfb}{{\bar \phi}{}^\beta}
\newcommand{\wpa}{\bar \pi_\alpha}
\newcommand{\wpb}{\bar \pi_\beta}

\newcommand{\fsam}{\phi^*_{\alpha,\mm}}
\newcommand{\fsbn}{\phi^*_{\beta,\nn}}
\newcommand{\psam}{\pi_*^{\alpha,\mm}}
\newcommand{\psbn}{\pi_*^{\beta,\nn}}
\newcommand{\Eam}{\EE_{\alpha,\mm}}

\newcommand{\wfsa}{\bar \phi^*_\alpha}
\newcommand{\wpsa}{\bar \pi_*^\alpha}
\newcommand{\wpsb}{\bar \pi_*^\beta}
\newcommand{\wfsam}{\bar \phi^*_{\alpha,\mm}}
\newcommand{\wfsan}{\bar \phi^*_{\alpha,\nn}}
\newcommand{\wfsbn}{\bar \phi^*_{\beta,\nn}}
\newcommand{\wpsam}{\bar \pi_*^{\alpha,\mm}}
\newcommand{\wpsan}{\bar \pi_*^{\alpha,\nn}}
\newcommand{\wpsbm}{\bar \pi_*^{\beta,\mm}}
\newcommand{\wpsbn}{\bar \pi_*^{\beta,\nn}}

\newcommand{\dam}{\d_\alpha^\mm}
\newcommand{\dbn}{\d_\beta^\nn}

\newcommand{\ord}{o}
\newcommand{\ordg}{\varsigma}

\newcommand{\Np}[1]{{N+p\choose N #1}}
\newcommand{\Npr}{{N+p-r\choose N-r}}
\newcommand{\ritwo}{(-)^i {r-2\choose i-2}}
\newcommand{\rione}{(-)^i {r-1\choose i-1}}

\newcommand{\si}{\sigma}
\newcommand{\eps}{\epsilon}
\newcommand{\dlt}{\delta}
\newcommand{\om}{\omega}
\newcommand{\al}{\alpha}
\newcommand{\bt}{\beta}
\newcommand{\gm}{\gamma}
\newcommand{\ka}{\kappa}
\newcommand{\la}{\lambda}
\newcommand{\vth}{\vartheta}
\renewcommand{\th}{\theta}
\newcommand{\rep}{\varrho}

\newcommand{\vect}{{\mathfrak{vect}}}
\newcommand{\map}{{\mathfrak{map}}}
\newcommand{\dmap}{\vect(N)\ltimes \map(N,\oj)}

\newcommand{\im}{{\rm im}\ }
\newcommand{\ext}{{\rm ext}\ }
\newcommand{\e}{{\rm e}}
\renewcommand{\div}{{\rm div}}
\newcommand{\afn}{{\rm afn\,}}
\newcommand{\til}{{\tilde{\ }}}

\newcommand{\eikx}{\e^{ik\cdot x}}
\newcommand{\dNx}{{d^N \kern-0.4mm x}}
\newcommand{\dNk}{{d^N \kern-0.4mm k}}
\newcommand{\dNxp}{{d^N \kern-0.4mm x'}}
\newcommand{\dNxb}{{d^N \kern-0.4mm x''}}
\newcommand{\dNkp}{{d^N \kern-0.4mm k'}}
\newcommand{\dFx}{{d^4 \kern-0.4mm x}}

\newcommand{\summ}[1]{\sum_{|\mm|\leq #1}}
\newcommand{\sumn}[1]{\sum_{|\nn|\leq #1}}
\newcommand{\summnp}{\sum_{|\mm|\leq|\nn|\leq p }}
\newcommand{\sumnmp}[1]{\sum_{|\nn|\leq|\mm|\leq p#1 }}
\newcommand{\summp}{\summ{p}}

\newcommand{\dmu}{{\d_\mu}}
\newcommand{\dnu}{{\d_\nu}}

\newcommand{\larroww}[1]{{\ \stackrel{#1}{\longleftarrow}\ }}
\newcommand{\rarroww}[1]{{\ \stackrel{#1}{\longrightarrow}\ }}
\newcommand{\intdm}{\sum_{m=-\infty}^\infty }

\newcommand{\repi}{\rep^{(i)}}
\newcommand{\Mi}{M^{(i)}}
\newcommand{\mri}{(-)^i {r\choose i}}

\newcommand{\mndmn}{{\mm\choose\nn}\d_{\mm-\nn}}

\newcommand{\trrep}{{\rm tr}_{\rep}\kern0.7mm}
\newcommand{\trM}{{\rm tr}_{M}\kern0.7mm}
\newcommand{\tr}{{\rm tr}}
\newcommand{\oj}{{\mathfrak g}}
\newcommand{\g}{{\mathfrak g}}
\newcommand{\hh}{{\mathfrak h}}
\newcommand{\uu}{{\mathfrak u}}

\renewcommand{\L}{{\cal L}}
\newcommand{\J}{{\cal J}}
\newcommand{\D}{{\cal D}}
\newcommand{\U}{{\cal U}}
\newcommand{\N}{{\cal N}}
\newcommand{\OO}{{\cal O}}
\newcommand{\QQ}{{\cal Q}}
\newcommand{\PP}{{\cal P}}
\newcommand{\EE}{{\cal E}}
\newcommand{\FF}{{\cal F}}
\newcommand{\HH}{{\cal H}}
\newcommand{\II}{{\cal I}}
\newcommand{\GG}{{\cal G}}

\newcommand{\cl}{{cl}}
\newcommand{\qm}{{qm}}

\newcommand{\TT}{{\mathbb T}}
\newcommand{\RR}{{\mathbb R}}
\newcommand{\CC}{{\mathbb C}}
\newcommand{\ZZ}{{\mathbb Z}}
\newcommand{\NN}{{\mathbb N}}

\title{{Manifestly Covariant Canonical Quantization of Gravity and
Diffeomorphism Anomalies in Four Dimensions}}

\author{T.A. Larsson \\
Vanadisv\"agen 29, S-113 23 Stockholm, Sweden\\
email: thomas.larsson@hdd.se}

\date{}

\maketitle

\begin{abstract}
Canonical quantization of gravity requires knowledge about the
representation theory of its constraint algebra, which is physically
equivalent to the algebra of arbitrary 4-diffeomorphisms. All interesting
lowest-energy representations are projective, making the relevant algebra
into a four-dimensional generalization of the Virasoro algebra. Such
diffeomorphism anomalies are invisible in field theory, because the
relevant cocycles are functionals of the observer's trajectory in
spacetime. The multi-dimensional Virasoro algebra acts naturally in the
phase space of arbitrary histories, with dynamics playing the role of
first-class constraints. General relativity is regularized by expanding
all fields in Taylor series around the observer's trajectory, and
truncating at some fixed order. This regularized but manifestly
general-covariant theory is quantized in the history phase space, and
dynamics is imposed afterwards, in analogy with BRST quantization.
Infinities arise when the regularization is removed; it is presently
unclear how these should be dealt with.
\end{abstract}

\medskip

\noindent
In: Focus on quantum gravity research	\hfill\break
ed: David C. Moore, pp 261-310		\hfill\break
2006 Nova Science Publishers, Inc.	\hfill\break
ISBN 1-59454-660-6			\hfill

\newpage

\section{Introduction}
\label{sec:Intro}

After its invention by Dirac in 1927, local relativistic quantum field
theory (QFT) has undergone three major crises.

The first crisis occurred in the 1930s, when it became clear that quantum
mechanics (QM) was na\"\i vely incompatible with special relativity, and
in particular that quantization of electromagnetism gave infinite
answers. This led many people to believe that
radical new physics was needed. Alas, the problems were solved by
renormalization, which involves no new physics, but merely a
reinterpretation (albeit a radical one) of old physics.

The second crisis occurred in the 1960s, when QFT appeared to be
incapable of describing the strong and weak interactions, in particular
asymptotic freedom. Therefore many people turned to
other, allegedly more fundamental ideas, such as analyticity of the
S-matrix or string theory. However, it again turned out that
straightforward QFT was the correct answer once it was applied to
non-abelian Yang-Mills theory.

The third crisis is of course the apparent incompatibility between QM and
gravity. This problem was realized already around 1960, when it was
found that perturbative quantum gravity is not renormalizable and hence
not a predictive theory, but it only became acute when the
standard model was completed 25 years ago, and gravity remained the only
interaction not described by a consistent QFT. As during the previous
crises, the difficulties with gravity has led many people to abandon QFT
in favor of more fashionable ideas.

The philosophy of this author is that quantum gravity should also be
described by straightforward QFT, but that a small but profound fix is
needed. The fix should be small, in order to preserve the experimental
successes of the standard model and general relativity, but it must also
be profound, because otherwise it would already have been found. This
philosophy is in many ways similar to renormalization, which can be
regarded as a small but profound fix which made QM compatible with
locality.

The key idea in the present work is that we need to quantize not only
the fields, but also the observer's trajectory in spacetime. In
non-relativistic QM, observation is a complicated, non-local process
which assigns numbers to experiments. However, in a relativistic theory,
a process must be localized; it happens {\em somewhere}. In order to
maintain locality, we must assign the process of observation to some
definite event in spacetime. As time proceeds, the observer (or detector
or test particle) traces out a curve $q(t)$ in spacetime. Like the
quantum fields, the observer's trajectory should be treated as a
material, quantized object; it has a conjugate momentum, it is
represented on the Hilbert space, etc.

To consider the observer's trajectory as a material object is certainly a
very small fix, which does not introduce any new physical ideas. One may
wonder why such a small, almost trivial modification should be important.
The reason is that it
makes it possible for new types of anomalies to arise. If the field
theory has a gauge symmetry of Yang-Mills type, there is a gauge anomaly
proportional to the quadratic Casimir, and the gauge (or current) algebra
becomes a higher-dimensional generalization of affine Kac-Moody algebras.
Similarly, a general-covariant theory, in any number of dimensions,
acquires a diffeomorphism anomaly, which is described by a
higher-dimensional generalization of the Virasoro algebra. The reason
why these anomalies can not be seen in conventional QFT, without explicit
reference to the observer, is that the relevant cocycles are functionals
of the observer's trajectory. If this trajectory has not been introduced,
it is of course impossible to write down the relevant anomalies.

At this point it is necessary to discuss the issue of gauge anomalies and
consistency, in particular unitarity. Gauge and diffeomorphism anomalies
are usually considered as a sign of inconsistency and should therefore be
cancelled \cite{Bon86,NAG85}.
There is ample evidence that this is the correct prescription for
conventional gauge anomalies, arising from chiral fermions coupled to
gauge fields. However, the anomalies discussed in this paper are of a
completely different type, depending on the observer's trajectory, and
intuition derived from conventional anomalies needs not apply. Instead,
the situation is similar to conformal field theory applied to
two-dimensional statistical physics, where it is well known that infinite
spacetime symmetry (gauge or not) is compatible with locality (in the
sense of correlation functions depending on separation) only in the
presence of a conformal anomaly. Hence gauge anomalies are viewed as a
means to gauge symmetry breaking.

Be that as it is. Even if we would want to mod out the new gauge and diff
anomalies (which this author believes is wrong), we still need to know
about them. A good analogy is bosonic string theory, whose conformal
anomaly cancels precisely in 26 dimensions. However, if we did not know
about conformal anomalies, there would be no condition that singled out
the number 26. Similarly, we need to know about the new anomalies that
arise when we quantize gauge-invariant or diff-invariant theories in the
presence of the observer's trajectory, even if all we wanted to do with
these anomalies were to cancel them.

Anomalies manifest themselves as extensions of the constraint algebras.
Extensions of the diffeomorphism algebra were classified by
Dzhumadildaev \cite{Dzhu96}, and their representation theory was
developed in
\cite{BB98,BBS01,Bil97,Bil03,Lar98,Lar01,Lar02,RM94};
see \cite{Lar03} for a recent review.
Unfortunately, the canonical formalism is not very well suited for
quantization of relativistic theories, because the foliation of spacetime
into fixed time slices breaks manifest covariance. This problem becomes
especially serious in general-covariant theories, where the very
constraint algebra is modified, from the algebra of arbitrary
4-diffeomorphisms into the Dirac algebra of constraints. However, this
modification is a consequence of the chosen formalism rather than a true
physical effect, and in covariant approaches the constraint algebra of
general relativity is indeed the 4-diffeomorphism algebra.

It is well known that phase space is a covariant concept; it is the space
of histories which solve the dyna\-mics. Each phase space point $(q,p)$
generates a unique history $(q(t),p(t))$ under Hamiltonian evolution, and
thus we may view $(q,p) = (q(0),p(0))$ as a particular coordinatization
of phase space. We can now understand how the Dirac algebra arises. A
canonical transformation in the history phase space maps the history
$(q(t),p(t)) \to (q'(t'),p'(t'))$. If in particular
\[
(q,p) = (q(0),p(0)) \rarroww{\hbox{symm}} (q'(0),p'(0)) = (q',p'),
\]
the transformation preserves the standard coordinatization; this is the
situation with spatial diffeomorphisms in general relativitiy. However,
temporal diffeomorphisms break the standard coordinatization, and if we
insist on keeping it, we must add a compensating transformation to move
back to the $t=0$ surface. Schematically,
\[
(q,p) = (q(0),p(0)) \rarroww{\hbox{symm}} (q'(t),p'(t))
\rarroww{\hbox{comp}} (q'(0),p'(0)) = (q',p').
\]
The combination of symmetry and compensating transformations generate
the Dirac algebra.

It is clear that we can avoid this complication if we work in the history
phase space directly, because then we do not need to worry about
compensating transformations. To this end, a novel quantization scheme
has recently been proposed, called manifest covariant canonical
quantization (MCCQ) \cite{Lar04,Lar05a}.
The idea is to make the space of arbitrary histories $(q(t),p(t))$ into
a phase space $\PP$ by defining the Poisson brackets
\[
[p(t), q(t')] = \dlt(t-t'), \qquad
[p(t), p(t')] = [q(t), q(t')] = 0.
\]
The Euler-Lagrange equations now define a constraint $\EE(t)\approx0$ in
$\PP$; since $\EE(t)$ only depends on $q(t)$ this constraint is first
class. This observation allows us to apply powerful cohomological methods
from BRST quantization of theories with first class constraints. In other
words, the idea in MCCQ is to quantize in the history phase space first
and to impose dyna\-mics afterwards, by passing to cohomology.
Since dyna\-mics is regarded as a constraint, the cohomology is
nontrivial even for systems without gauge symmetries, like the harmonic
oscillator and the free scalar field.

It is important to understand which ideas are crucial and which are
merely convenient. The formulation of QFT using MCCQ is ``just
formalism''. It is very convenient to have a canonical quantization
scheme which respects general covariance, but it is probably possible to
reexpress the results in this paper using non-covariant canonical
quantization, at the price of great complications. In contrast, adding
the observer's trajectory to the quantum fields is absolutely essential.
Without it, the new observer-dependent anomalies can not be formulated,
and this is a hard obstruction to quantization.

History methods have recently been advocated by Savvidou and Isham
\cite{Ish95,Sav99,Sav04}; in particular, the last reference contains a
very good summary of the conceptual problems involved in non-covariant
canonical quantization. Their formalism differs in details from MCCQ,
e.g. because they do not use cohomological methods. There is also a
substantial difference, namely that the observer's trajectory is not
introduced, and hence no diffeomorphism anomalies are seen.

Finally, let us emphasize the philosophical motivation for introducing
MCCQ. The representations considered here are of lowest-energy type,
i.e. there is a natural Hamiltonian whose eigenvalues are bounded from
below. This is the kind of representations expected to be relevant to
quantum theory. However, they do not look like standard formulations of
QFT, because their natural habitat is in history space. To apply them to
physics, we must first recast physics in a suitable, history-oriented
form. It is the same argument that leads us to use tensor calculus in
general relativity; the classical irreps of the diffeomorphism group
act on modules of tensor fields, so we should formulate physics in
terms of those.

This article is organized as follows.

In the next section, the relation between anomalies, consistency,
locality and unitarity is further discussed, using the infinite conformal
symmetry in two-dimensional spacetime as a paradigm.
The new anomalies, i.e. the higher-dimensional generalizations of the
affine and Virasoro algebras, are reviewed in Section \ref{sec:Virasoro}.
For easy comparison to the one-dimensional case, we describe these
extensions in a Fourier basis on the $N$-dimensional torus.
However, the geometrical content is clearer in a real-space basis,
which is introduced in Section \ref{sec:DGRO}. It turns out that in
addition to the diffeomorphism algebra $\vect(N)$ (algebra of vector
fields in $N$ dimensions) and the gauge algebra $\map(N,\oj)$ (algebra
of maps from $N$-dimensional spacetime to the finite-dimensional Lie
algebra $\g$), we must also introduce the observer's trajectory and
the algebra of repara\-metrizations; the full algebra is called
the DGRO (Diffeomorphism, Gauge, Repara\-metri\-zation, Observer)
algebra $DGRO(N,\oj)$ .

Its representation theory is developed in Section \ref{sec:DGROrep}.
Rather than starting from the fields themselves, as is done in one
dimension, the right approach is to first expand all fields in a Taylor
series around the observer's trajectory and truncate at some finite order
$p$ before quantization. This gives us a non-linear realization of the
diffeomorphism algebra on finitely many functions of a {\em single}
variable, which is precisely the situation where normal ordering works.
We also get an action of a $\vect(1)$ algebra describing
repara\-metrizations for free, i.e. without enlarging the realization.

In Section \ref{sec:MCCQ} we introduce the manifestly covariant canonical
quantization scheme mentioned above, and apply it to the free scalar
field in Section \ref{sec:Scalar1}. However, the Hamiltonian still
singles out a preferred time direction. To remedy this, we introduce in
Section \ref{sec:MCCQ-jets} the observer and define the Hamiltonian
covariantly as the operator which translates the fields relative to the
observer. The scalar field is again used as an example in
Section \ref{sec:Scalar2}. The formalism is then extended to theories
with gauge symmetries in Section \ref{sec:MCCQ-gauge}.
As examples we treat the free Maxwell field in Section \ref{sec:Maxwell}
and pure Einstein gravity in Section \ref{sec:Gravity}. The $\uu(1)$ gauge
symmetry of the Maxwell field is anomaly free in the absense of matter,
but there are diffeomorphism anomalies already in pure gravity, because
it is an interacting theory.

Truncating the Taylor expansion at order $p$ is a regularization, and at
the end we want to remove the regulator by taking the limit $p\to\infty$.
This limit is problematic and poorly understood, as is discussed in
Section \ref{sec:Finiteness}. We conclude with a conceptual discussion
in Sections \ref{sec:Concept} and \ref{sec:Conclusion}.

\section{Anomalies, consistency, locality, and unitarity}
\label{sec:Locality}

A quantum theory is defined by a Hilbert space and a Hamiltonian which
generates time evolution. The main conditions for consistency are
unitarity and lack of infinities. If the theory has some symmetries,
these must be realized as unitary operators acting on the Hilbert space
as well. In particular, if time translation is included among the
symmetries, which is the case for the Poincar\'e and diffeomorphism
algebras, a unitary representation of the symmetry algebra is usually
enough for consistency. {F}rom this viewpoint, there is a 1-1
correspondence between general-covariant QFTs and unitary
representations of the diffeomorphism group on a conventional Hilbert
space. Namely, given the QFT, its Hilbert space carries a unitary
representation of the diffeomorphism group. Conversely, if we have a
unitary representation of the diffeomorphism group, the Hilbert space on
which it acts can be interpreted as the Hilbert space of some
general-covariant QFT.

Unfortunately, this observation is not yet so powerful, because no
non-trivial, unitary, lowest-energy irreps of the diffeomorphism algebra
are known except in one dimension. Nevertheless, we are able to make
some very general observations.
Assume that some algebra (or group) $\g$ has a unitary representation $R$
and a subalgebra $\hh$. Then the restriction of $R$ to $\hh$ is still
unitary, and this must hold for every subalgebra $\hh$ of $\g$. In
particular, let $\g = \vect(N)$ be the diffeomorphism algebra in $N$
dimensions and $\hh = \vect(1)$ the diffeomorphism algebra in one
dimension. There are infinitely many such subalgebras, and the restriction
of $R$ to each and every one of them must be unitary. Fortunately, the
unitary irreps of the diffeomorphism algebra in one dimension are known.
The result is that the only proper unitary irrep is the trivial one, but
there are many unitary irreps with a diffeomorphism anomaly. {F}rom this
it follows that the trivial representation is the only unitary
representation also in $N$ dimensions.``There are no local observables
in quantum gravity''.

However, there is one well-known case where we know how to combine
locality and infinite spacetime symmetry with quantum theory: conformal
field theory (CFT). This is usually thought of as a theory of conformally
invariant quantum fields in two dimensions, but since the local
conformal group is
the same as (twice) the diffeomorphism group in one dimension, it is also
about diffeomorphism invariant QFT in one dimension. Locality means that
the correlation functions depend on separation. For two points $z$ and $w$
in $\RR$ or $\CC$, the correlator is
\be
  \langle \phi(z) \phi(w) \rangle \sim {1\/(z-w)^{2h}} + more,
\label{correlator}
\ee
where $more$ stands for less singular terms when $z \to w$. That the
correlation function has this form is a diffeomorphism-invariant
statement. The
$more$ terms will change under an arbitary diffeomorphism, but the
leading singularity will always have the same form, and in particular
the anomalous dimension $h$ is well defined.

We can phrase this slightly differently. The short-distance singularity
only depends on two points being infinitesimally close. This is good,
because we cannot determine the finite distance between two points
without knowing about the metric. General relativity does not have a
background metric structure, but it does have a background differentiable
structure (locally at least), and that is enough for defining anomalous
dimensions. Diffemorphisms move points around, but they do not separate
two points which are infinitesimally close.

The relevant algebra in CFT is not really the one-dimensional
diffeomorphism algebra (or the two-dimensional conformal algebra), but
rather its central extension known as the Virasoro algebra:
\be
  [L_m, L_n] = (n-m) L_{m+n} - {c\/12} (m^3 - m) \delta_{m+n}.
\ee
A lowest-energy representation is characterized by a vacuum satisfying
\be
  L_0 \ket{0} = h \ket{0},
  \qquad L_m \ket{0} = 0\ \hbox{for all $m < 0$.}
\ee
In particular, the lowest $L_0$ eigenvalue can be identified with the
anomalous dimension $h$ in the correlation function (\ref{correlator}).
This means that locality, in the sense of correlation functions
depending on separation, requires that $h > 0$. It is well known
\cite{FMS96} that unitarity either implies that
\be
c = 1 - {6\/m(m+1)}, \qquad
h = h_{rs}(m) = {[(m+1)r-ms]^2-1\/4m(m+1)},
\ee
where $m\geq2$ and $1\leq r<m, 1\leq s<r$ are positive integers,
or that $c\geq1$, $h\geq0$. In particular, the central extension $c$ is
non-zero for any non-trivial, unitary irrep with $h \neq 0$. This leads
to the important observation:

\bigskip
\begin{tabular}{|l|}
\hline
Locality and unitarity are compatible with diffeomorphism (and \\
local conformal) symmetry only in the presence of an anomaly.  \\
\hline
\end{tabular}
\bigskip

This is true in higher dimensions as well. Consider the correlator
$\langle\phi(x) \phi(y)\rangle$, where $x$ and $y$ are points in $\RR^N$.
We could take some one-dimensional curve $q(t)$ passing through $x$ and
$y$, such that $x = q(t)$ and $y = q(t')$. Then the short-distance
behaviour is of the form
\be
  \langle \phi(x) \phi(y) \rangle \sim {1\/(t - t')^{2h}} + more,
\ee
and $h$ is independent of the choice of curve, provided that it is
sufficiently regular. The subalgebra of $\vect(N)$ which preserves
$q(t)$ is a Virasoro algebra, so $h > 0$ implies that $c > 0$.

This observation is completely standard in the application of CFT to
statistical physics in two dimensions. The simplest example of a unitary
model is the Ising model, which consists of three irreps, with $c = 1/2$
and $h = 0$, $h = 1/16$, and $h = 1/2$. The Ising model is perfectly
consistent despite the anomaly, both mathematically (unitarity), and more
importantly physically (it is realized in nature, in soft condensed
matter systems). The standard counter-argument is that infinite conformal
symmetry in condensed matter is not a gauge symmetry, but rather an
anomalous global symmetry. However, if we could take the classical limit
of such a system, the conformal symmetry would seem to be a gauge
symmetry. Namely, the anomaly vanishes in the classical limit, and we can
write down a classical BRST operator which is nilpotent, and the symmetry
is gauge on the classical level. There is no classical way to distinguish
between such a ``fake'' gauge symmetry and a genuine gauge symmetry which
extends to the quantum level.

More generally, let us assume that we have some phase space, and a Lie
algebra $\g$ with generators $J_a$, satisfying
\be
  [J_a, J_b] = f_{ab}{}^c J_c,
\label{oj}
\ee
acts on this phase space. The Einstein convention is used;
repeated indices, one up and one down, are implicitly summed over.
If the bracket with the Hamiltonian gives us
a new element in $\g$,
\be
  [J_a, H] = C^b_a J_b,
\ee
we say that $\g$ is a {\em symmetry} of the Hamiltonian system. If $\g$ in
addition contains arbitary functions of time, the symmetry is a {\em gauge
symmetry}. In this case, a solution to Hamilton's equations depends on
arbitrary functions of time and is thus not fully specified by the
positions and momenta at time $t = 0$. The standard example is
electromagnetism, where the zeroth component $A_0$ of the vector
potential is arbitrary, because its canonical momentum $F^{00} = 0$. An
arbitary time evolution is of course not acceptable. The reason why
this seems to happen is that a gauge symmetry is a redundancy of the
description; the true dynamical degrees of freedom are fewer than what
one na\"\i vely expects. In electromagnetism, the gauge potential has
four components but the photon has only two polarizations.

There are various ways to handle quantization of gauge systems. One is to
eliminate the gauge degrees of freedom first and then quantize. This is
cumbersome and it is usually preferable to quantize first and eliminate
the gauge symmetries afterwards. The simplest way
is to require that the gauge generators annihilate physical states,
\be
  J_a \ket{phys} = 0,
\ee
and also that two physical states are equivalent if the differ by some
gauge state, $J_a |>$. This procedure produces a Hilbert space of
physical states.

However, one thing may go wrong. Upon quantization, a symmetry may
acquire some quantum corrections, so that $\g$ is replaced by
\be
  [J_a, J_b] = f_{ab}{}^c J_c + \hbar D_{ab} + O(\hbar^2).
\ee
The operator $D_{ab}$ is called an {\em anomaly}. We can also have
anomalies of the type
\be
  [J_a, H] = C_a^b J_b + \hbar E_a + O(\hbar^2).
\ee
If we now try to keep the definition of a physical state, we see that
we must also demand that
\be
  D_{ab} \ket{phys} = 0.
\ee
This implies further reduction of the Hilbert space. In the case that
$D_{ab}$ is invertible, there are no physical states at all,
so the Hilbert space is empty. However, this
does not necessarily mean that the anomaly by itself is inconsistent,
only that our definition of physical states is. In the presence of an
anomaly, additional states become physical. So our Hilbert space
becomes larger, containing some, or even all, of the previous gauge
degrees of freedom. A gauge anomaly implies that the gauge symmetry is
broken on the quantum level.

Such a ``fake'' gauge symmetry may well
be consistent. The Virasoro algebra is obviously anomalous, with the
central charge playing the role of the $D_{ab}$, and still it has unitary
representations with non-zero $c$. Of course, a gauge anomaly {\em may}
be inconsistent, if the anomalous algebra does not possess any unitary
representations. This is apparently what happens for the chiral-fermion
type anomaly which is relevant e.g. in the standard model.

\section{Multi-dimensional Virasoro algebra}
\label{sec:Virasoro}

All non-trivial, unitary, lowest-energy irreps of the diffeomorphism
algebra are anomalous, in any number of dimensions. This is well-known in
one dimension, where the diffeomorphism algebra acquires an extension
known as the Virasoro algebra. It is also true in several dimensions,
which one proves by considering the restriction to the many Virasoro
subalgebras living on lines in spacetime. This is perhaps rather
surprising, in view of the following two no-go theorems:
\begin{itemize}
\item
The diffeomorphism algebra has no central extension except in one
dimension.
\item
In field theory, there are no pure gravitational anomalies in four
dimension.
\end{itemize}
However, the assumptions in these no-go theorems are too strong; the
Virasoro extension is not central except in one dimension, and one needs
to go slightly beyond field theory by explicitly specifying where
observation takes place.

To make contact with the Virasoro algebra in its most familiar form, we
describe its multi-dimensional sibling in a Fourier basis on the
$N$-dimensional torus. Recall first that the algebra of diffeomorphisms on
the circle, $\vect(1)$, has generators
\be
L_m = -i \exp(imx) {d\/d x},
\label{Lm}
\ee
where $x \in S^1$.
$\vect(1)$ has a central extension, known as the Virasoro algebra:
\be
[L_m, L_n] = (n-m)L_{m+n} - {c\/12} (m^3-m) \delta_{m+n},
\label{Vir}
\ee
where $c$ is a c-number known as the central charge or conformal
anomaly. This means that the Virasoro algebra is a Lie
algebra; anti-symmetry and the Jacobi identities still hold. The
term linear in $m$ is unimportant, because it can be removed by a
redefinition of $L_0$. The cubic term $m^3$ is a non-trivial
extension which cannot be removed by any redefinition.

The generators (\ref{Lm}) immediately generalize to vector fields on the
$N$-dimensional torus:
\be
L_\mu(m) = -i \exp(i m_\rho x^\rho) \dmu,
\ee
where $x = (x^\mu)$, $\mu = 1, 2, ..., N$ is a point in
$N$-dimensional space and $m = (m_\mu) \in \ZZ^N$.
These operators generate the algebra $\vect(N)$:
\be
[L_\mu(m), L_\nu(n)] = n_\mu L_\nu(m+n) - m_\nu L_\mu(m+n).
\ee
The question is now whether the Virasoro extension, i.e. the
$m^3$ term in (\ref{Vir}), also generalizes to higher dimensions.

Rewrite the ordinary Virasoro algebra (\ref{Vir}) as
\bes
[L_m, L_n] &=& (n-m)L_{m+n} + c m^2 n S_{m+n}, \nl
{[}L_m, S_n] &=& (n+m)S_{m+n}, \nle
{[}S_m, S_n] &=& 0, \nl
m S_m &\equiv& 0.
\eens
It is easy to see that the two formulations of the Virasoro algebra are
equivalent (the linear cocycle has been absorbed into a redefinition of
$L_0$). The second formulation immediately generalizes to $N$
dimensions. The defining relations are
\bes
[L_\mu(m), L_\nu(n)] &=& n_\mu L_\nu(m+n) - m_\nu L_\mu(m+n) \nl
&&  + (c_1 m_\nu n_\mu + c_2 m_\mu n_\nu) m_\rho S^\rho(m+n), \nl
{[}L_\mu(m), S^\nu(n)] &=& n_\mu S^\nu(m+n)
 + \delta^\nu_\mu m_\rho S^\rho(m+n),
\nlb{mVir}
{[}S^\mu(m), S^\nu(n)] &=& 0, \nl
m_\mu S^\mu(m) &\equiv& 0.
\eens
This is an extension of $\vect(N)$ by the abelian ideal with basis
$S^\mu(m)$.
Geometrically, we can think of $L_\mu(m)$ as a vector field
and $S^\mu(m) = \eps^{\mu\nu_2..\nu_N} \ab S_{\nu_2..\nu_N}(m)$
as a dual one-form (and $S_{\nu_2..\nu_N}(m)$ as an $(N-1)$-form);
the last condition expresses closedness.
The cocycle proportional to $c_1$ was discovered by
Rao and Moody \cite{RM94}, and the one proportional to $c_2$ by
this author \cite{Lar91}.

There is also a similar multi-dimensional generalization of affine
Kac-Moody algebras, presumably first written down by Kassel
\cite{Kas85}. It is sometimes called the central extension, but this
term is somewhat misleading because the extension does not commute with
diffeomorphisms, although it does commute with all gauge
transformations.

Let $\oj$ be a finite-dimensional Lie algebra with structure
constants $f_{ab}{}^c$ and Killing metric $\delta_{ab}$. The Kassel
extension of the current algebra $\map(N,\oj)$ is defined by the
brackets
\bes
[J_a(m), J_b(n)] &=& f_{ab}{}^c J_c(m+n)
  + k \delta_{ab} m_\rho S^\rho(m+n), \nl
{[}J_a(m), S^\mu(n)] &=& [S^\mu(m), S^\nu(n)] = 0,
\label{affine}\\
m_\mu S^\mu(m) &\equiv& 0.
\eens
This algebra admits an intertwining action of the $N$-dimensional
Virasoro algebra (\ref{mVir}):
\be
[L_\mu(m), J_a(n)] = n_\mu J_a(m+n).
\ee

The current algebra $\map(N,\oj)$ also admits another type of
extension in some dimensions. The best known example is the
Mickelsson-Faddeev algebra, relevant for the conventional
anomalies in field theory, which arise when chiral fermions are
coupled to gauge fields in three spatial dimensions. Let
$d_{abc} = \tr \{T_a, T_b\}T_c$ be the totally symmetric third
Casimir operator, and let $\eps^{\mu\nu\rho}$ be the totally
anti-symmetric epsilon tensor in three dimensions. The
Mickelsson-Faddeev algebra \cite{Mi89} reads in a Fourier basis:
\bes
[J_a(m), J_b(n)] &=& f_{ab}{}^c J_c(m+n)
  + d_{abc} \epsilon^{\mu\nu\rho} m_\mu n_\nu A^c_\rho(m+n), \nl
{[}J_a(m), A^b_\nu(n)] &=& -f_{ac}{}^b A^c_\nu(m+n)
  + \delta_a^b m_\nu \dlt(m+n),
\label{MF}\\
{[}A^a_\mu(m), A^b_\nu(n)] &=& 0.
\eens
$A^a_\mu(m)$ are the Fourier components of the gauge connection.

Note that $Q_a \equiv J_a(0)$ generates a Lie algebra isomorphic to
$\oj$, whose Cartan subalgebra is identified with the charges.
Moreover, the subalgebra of (\ref{affine}) spanned by
$J_a(m_0) \equiv J_a(m)$, where $m = (m_0, 0, ..., 0) \in \ZZ$, reads
\be
[J_a(m_0), J_b(n_0)] &=& f_{ab}{}^c J_c(m_0+n_0)
+ k \delta_{ab} m_0 \dlt(m_0+n_0),
\ee
which we recognize as the affine algebra $\widehat{\oj}$. Since all
non-trivial unitary irreps of $\widehat{\oj}$ has $k>0$ \cite{GO86},
it is impossible to combine unitary and non-zero $\oj$ charges also
for the higher-dimensional algebra (\ref{affine}).
This follows immediately from the fact that the restriction of a
unitary irrep to a subalgebra is also unitary (albeit in general
reducible).

In contrast, the Mickelsson-Faddeev algebra (\ref{MF}) has apparently no
faithful unitary representations on a separable Hilbert space
\cite{Pic89}. A simple way to understand this is to note that the
restriction to every loop subalgebra is proper and hence lacks unitary
representations of lowest-weight type. This presumably means that this
kind of extension should be avoided. Indeed, Nature appears to abhor this
kind of anomaly, which is proportional to the third Casimir.

\section{DGRO algebra}
\label{sec:DGRO}

The Fourier formalism in the previous section makes the analogy with
the usual Virasoro algebra manifest, but it is neither illuminating
nor a useful starting point for representation theory. To bring out
the geometrical content, we introduce the DGRO
{\em (Diffeomorphism, Gauge, Repara\-metri\-zation, Observer)} algebra
$DGRO(N,\oj)$, whose ingredients are spacetime diffeomorphisms which
generate $\vect(N)$,
repara\-metri\-zations of the observer's trajectory which form an
additional $\vect(1)$ algebra, and gauge transformations which generate
a current algebra. Classically, the algebra is $\dmap\oplus\vect(1)$.

Let $\xi=\xmu(x)\dmu$, $x\in\RR^N$, $\dmu = \d/\d x^\mu$,
be a vector field, with commutator
$[\xi,\eta] \equiv \xmu\dmu\ynu\dnu - \ynu\dnu\xmu\dmu$,
and greek indices $\mu,\nu = 1,2,..,N$ label the
spacetime coordinates.
The Lie derivatives $\L_\xi$ are the generators of $\vect(N)$.

Let $f = f(t)d/dt$, $t\in S^1$, be a vector field in one dimension.
The commutator reads
$[f,g] = (f\dot g - g\dot f)d/dt$, where the dot denotes the $t$
derivative: $\dot f \equiv df/dt$. We will also use $\d_t = \d/\d t$
for the partial $t$ derivative.
The choice that $t$ lies on the circle is physically unnatural and is
made for technical simplicity only (quantities can be expanded in
Fourier series). However, this seems to be a
minor problem at the present level of understanding.
Denote the repara\-metri\-zation generators $L_f$.

Let $\map(N,\oj)$ be the current algebra corresponding to the
finite-dimen\-sional semisimple Lie algebra $\oj$ with basis $J_a$,
structure constants $f_{ab}{}^c$, and Killing metric $\dlt_{ab}$.
The brackets in $\oj$ are given by (\ref{oj}).
A basis for $\map(N,\oj)$ is given by $\oj$-valued functions $X=X^a(x)J_a$
with commutator $[X,Y]=f_{ab}{}^c X^aY^bJ_c$. The intertwining
$\vect(N)$ action is given by $\xi X = \xmu\dmu X^a J_a$. Denote the
$\map(N,\oj)$ generators by $\J_X$.

Finally, let $Obs(N)$ be the space of local functionals of the
observer's trajectory $q^\mu(t)$, i.e. polynomial functions of
$q^\mu(t)$, $\dot q^\mu(t)$, ... $d^k  q^\mu(t)/dt^k$,  $k$ finite,
regarded as a commutative algebra. $Obs(N)$ is a $\vect(N)$ module in a
natural manner.

$DGRO(N,\oj)$ is an abelian but non-central Lie algebra extension of
$\vect(N) \ltimes \map(N,\oj) \oplus \vect(1)$ by $Obs(N)$:
\[
0 \longrightarrow Obs(N) \longrightarrow DGRO(N,\oj) \longrightarrow
 \vect(N)\ltimes \map(N,\oj)\oplus \vect(1) \longrightarrow 0.
\]
The brackets are given by
\bes
[\L_\xi,\L_\eta] &=& \L_{[\xi,\eta]}
 + {1\/2\pi i}\int dt\ \dot q^\rho(t)
 \Big\{ c_1 \d_\rho\dnu\xmu(q(t))\dmu\ynu(q(t)) +\nl
&&\quad+ c_2 \d_\rho\dmu\xmu(q(t))\dnu\ynu(q(t)) \Big\}, \nl
{[}\L_\xi, \J_X] &=& \J_{\xi X}, \nl
{[}\J_X, \J_Y] &=& \J_{[X,Y]} - {c_5\/2\pi i}\dlt_{ab}
 \int dt\ \dot q^\rho(t)\d_\rho X^a(q(t))Y^b(q(t)), \nl
{[}L_f, \L_\xi] &=& {c_3\/4\pi i} \int dt\
 (\ddot f(t) - i\dot f(t))\dmu\xmu(q(t)),
\label{DGRO}\\
{[}L_f,\J_X] &=& 0, \nl
{[}L_f,L_g] &=& L_{[f,g]}
 + {c_4\/24\pi i}\int dt (\ddot f(t) \dot g(t) - \dot f(t) g(t)), \nl
{[}\L_\xi, q^\mu(t)] &=& \xmu(q(t)), \nl
{[}L_f, q^\mu(t)] &=& -f(t)\dot q^\mu(t), \nl
{[}\J_X,  q^\mu(t)] &=& {[} q^\mu(s),  q^\nu(t)] = 0,
\eens
extended to all of $Obs(N)$ by Leibniz' rule and linearity.
The numbers $c_1-c_5$ are called {\em abelian charges}, in analogy with
the central charge of the Virasoro algebra. In \cite{Lar98,Lar01}
slightly more complicated extensions were considered, which
depend on three additional abelian charges $c_6-c_8$. However, these
vanish automatically when $\oj$ is semisimple.

\section{Representations of the DGRO algebra}
\label{sec:DGROrep}

To construct Fock representations of the ordinary Virasoro algebra is
straightforward:
\begin{itemize}
\item
Start from classical modules, i.e. primary fields = scalar densities.
\item
Introduce canonical momenta.
\item
Normal order.
\end{itemize}
The first two steps of this procedure generalize nicely to higher
dimensions. The classical representations of the DGRO algebra are tensor
fields over $\RR^N \times S^1$ valued in $\oj$ modules. The basis of a
classical DGRO module $\QQ$ is thus a field
$\fa(x,t)$, $x\in\RR^N$, $t\in S^1$,
where $\al$ is a collection of all kinds of indices.
The $DGRO(N,\oj)$ action on $\QQ$ can be succinctly summarized as
\bes
[\L_\xi, \fa(x,t)] &=& -\xmu(x)\dmu\fa(x,t)
- \dnu\xmu(x)T^{\al\nu}_{\bt\mu}\fb(x,t), \nl
{[}\J_X, \fa(x,t)] &=& -X^a(x)J^{\al}_{\bt a}\fb(x,t),
\\
{[}L_f, \fa(x,t)] &=& -f(t)\d_t\fa(x,t)
- \la(\dot f(t)-if(t))\fa(x,t).
\eens
Here $J_a = (J^\al_{\bt a})$ and $T^\mu_\nu = (T^{\al\mu}_{\bt\nu})$ are
matrices satisfying $\oj$ (\ref{oj}) and $gl(N)$, respectively:
\be
[T^\mu_\nu,T^\si_\tau] =
\dlt^\si_\nu T^\mu_\tau - \dlt^\mu_\tau T^\si_\mu.
\label{glN}
\ee
The tensor field representations of the DGRO algebra can thus be
expressed in matrix form as
\bes
\L_\xi &=& -\int \dNx\int dt\
(\xmu(x)\dmu\fa(x,t) +\dnu\xmu(x)T^{\al\nu}_{\bt\mu}\fb(x,t))
\pa(x,t), \nl
\J_X &=& -\int \dNx\int dt\
X^a(x)J^{\al}_{\bt a}\fb(x,t) \pa(x,t),
\label{class} \\
L_f &=& -\int \dNx \int dt\
(f(t)\d_t\fa(x,t) - \la(\dot f(t)-if(t))\fa(x,t)) \pa(x,t),
\eens
where the conjugate momentum $\pa(x,t) = \dd{\fa(x,t)}$ satisfies
\be
[\pa(x,t), \fb(x',t')] = \dlt^\bt_\al \dlt(x-x') \dlt(t-t').
\ee

However, the normal-ordering step simply does not work in several
dimensions, because
\begin{itemize}
\item
It requires that a foliation of spacetime into space and time has been
introduced, which runs against the idea of diffeomorphism invariance.
\item
Normal ordering of bilinear expressions always results in a {\em central}
extension, but the Virasoro cocycle is non-central when $N\geq 2$.
\item
It is ill defined. Formally, attempts to normal order result in an
{\em infinite} central extension, which of course makes no sense.
\end{itemize}
To avoid this problem,
the crucial idea in \cite{Lar98} was to expand all fields in a Taylor
series around the observer's trajectory and truncate at order $p$,
before introducing canonical momenta.
Hence we expand e.g.,
\be
\fa(x,t) = \summ p {1\/\mm!} \fam(t)(x-q(t))^\mm,
\label{Taylor}
\ee
where $\mm = (m_1, \ab m_2, \ab ..., \ab m_N)$, all $m_\mu\geq0$, is a
multi-index of length $|\mm| = \sum_{\mu=1}^N m_\mu$,
$\mm! = m_1!m_2!...m_N!$, and
\be
(x-q(t))^\mm = (x^1-q^1(t))^{m_1} (x^2-q^2(t))^{m_2} ...
 (x^N-q^N(t))^{m_N}.
\label{power}
\ee
Denote by $\mu$ a unit vector in the $\mu$:th direction, so that
$\mm+\mu = (m_1, \ab ...,m_\mu+1, \ab ..., \ab m_N)$, and let
\be
\fam(t) = \d_\mm\fa(q(t),t)
= \underbrace{\d_1 .. \d_1}_{m_1} ..
\underbrace{\d_N .. \d_N}_{m_N} \fa(q(t),t)
\label{jetdef}
\ee
be the $|\mm|$:th order derivative of $\fa(x,t)$ evaluated on the
observer's trajectory $q^\mu(t)$.

Given two jets $\phi_\cmm(t)$ and $\psi_\cmm(t')$, we define their product
\be
(\phi(t)\psi(t'))_\cmm
= \sum_\nn {\mm\choose\nn}\phi_\cnn(t) \psi_{,\mm-\nn}(t').
\label{product}
\ee
It is clear that $(\phi(t)\psi(t'))_\cmm$ is the jet corresponding to the
field $\phi(x,t)\psi(x,t')$. For brevity, we also denote
$(\phi\psi)_\cmm(t) = (\phi(t)\psi(t))_\cmm$.

$p$-jets transform under $DGRO(N,\oj)$ as
\bes
[\L_\xi, \fam(t)] &=& \d_\mm([\L_\xi,\fa(q(t),t)])
+ [\L_\xi, q^\mu(t)]\dmu\d_\mm\fa(q(t),t) \nl
&\equiv& -\sumnmp{} T^{\al\nn}_{\bt\mm}(\xi(q(t))) \fbn(t), \nl
{[}\J_X, \fam(t)] &=& \d_\mm([\J_X,\fa(q(t),t)])
\label{jet} \\
&\equiv& -\sumnmp{} J^{\al\nn}_{\bt\mm}(X(q(t))) \fbn(t), \nl
{[}L_f, \fam(t)] &=& -f(t)\dotfam(t)
- \la(\dot f(t)-if(t))\fam(t),
\eens
where
\bes
T^\mm_\nn(\xi) &\equiv& (T^{\al\mm}_{\bt\nn}(\xi))
= \sum_{\mu\nu}{\nn\choose\mm} \d_{\nn-\mm+\nu}\xmu T^\nu_\mu \nl
&&\qquad + \sum_\mu{\nn\choose\mm-\mu}\d_{\nn-\mm+\mu}\xmu
  - \sum_\mu \dlt^{\mm-\mu}_\nn \xmu,
\label{Tmn}\\
J^\mm_\nn(X) &\equiv& (J^{\al\mm}_{\bt\nn}(X))
= {\nn\choose\mm} \d_{\nn-\mm} X^a J_a,
\eens
and
\be
{\mm\choose\nn} = {\mm!\/\nn!(\mm-\nn)!} =
{m_1\choose n_1}{m_2\choose n_2}...{m_N\choose n_N}.
\ee

We thus obtain a non-linear realization of $\vect(N)$ on the space of
trajectories in the space of tensor-valued
$p$-jets\footnote{$p$-jets are usually defined
as an equivalence class of functions: two functions are equivalent if all
derivatives up to order $p$, evaluated at $q^\mu$, agree. However, each
class has a unique representative which is a polynomial of order at most
$p$, namely the Taylor expansion around $q^\mu$, so we may canonically
identify jets with truncated Taylor series. Since $q^\mu(t)$ depends on a
parameter $t$, we deal in fact with trajectories in jet space, but these
will also be called jets for brevity.};
denote this space by $J^p\QQ$. Note that $J^p\QQ$ is spanned by $q^\mu(t)$
and $\{\fam(t)\}_{|\mm|\leq p}$ and thus not a $DGRO(N,\oj)$ module by
itself, because diffeomorphisms act non-linearly on $q^\mu(t)$, as can
be seen in (\ref{DGRO}). However, the space $C(J^p\QQ)$ of functionals on
$J^p\QQ$ (local in $t$) {\em is} a module, because the action on a $p$-jet
can never produce a jet of order higher than $p$. The space
$C(q) \otimes_q J^p\QQ$, where only the trajectory itself appears
non-linearly, is a submodule.

The crucial observation is that the jet space $J^p\QQ$
consists of finitely many functions of a single
variable $t$, which is precisely the situation where the normal ordering
prescription works. After normal ordering, denoted by double dots $:\ :$,
we obtain a Fock representation of the DGRO algebra:
\bes
\L_\xi &=& \int dt\ \Big\{ \no{\xmu(q(t))  p_\mu(t)} -
\sum_{|\nn|\leq|\mm|\leq p}
T^{\al\nn}_{\bt\mm}(\xi(q(t))) \no{ \fbn(t)\pam(t) } \Big\}, \nl
\J_X &=& -\int dt\ \Big\{ \sum_{|\nn|\leq|\mm|\leq p}
J^{\al\nn}_{\bt\mm}(\xi(q(t))) \no{ \fbn(t)\pam(t) } \Big\},
\label{Fock}\\
L_f &=& \int dt\ \Big\{ -f(t)\no{\dotfam(t)\pam(t)}
- \la(\dot f(t)-if(t))\no{\fam(t)\pam(t)} \Big\},
\eens
where we have introduced canonical momenta
$p_\mu(t) = \dd{ q^\mu(t)}$ and $\pam(t) = \dd{\fam(t)}$.
The field $\fa(x,t)$ can be either bosonic or fermionic but the
trajectory $q^\mu(t)$ is of course always bosonic.

Normal ordering is defined with respect to frequency; any function of
$t \in S^1$ can be expanded in a Fourier series, e.g.
\be
 p_\mu(t) = \intdm \hat p_\mu(m) \e^{-imt} \equiv
 p_\mu^<(t) + \hat p_\mu(0) +  p_\mu^>(t),
\label{Fourier}
\ee
where $p_\mu^<(t)$ ($p_\mu^>(t)$) is the sum over negative (positive)
frequency modes only. Then
\be
\no{\xmu(q(t))  p_\mu(t)}
\equiv \xmu(q(t))  p_\mu^<(t) +  p_\mu^>(t) \xmu(q(t)),
\ee
where the zero mode has been included in $p_\mu^<(t)$.

It is clear that (\ref{Fock}) defines a Fock representation for every
$gl(N)$ irrep $\rep$ and every $\oj$ irrep $M$; denote this Fock space
by $J^p\FF$, which indicates that it also depends on the truncation
order $p$. Namely, introduce a
Fock vacuum $\ket0$ which is annihilated by half of the oscillators,
i.e.
\be
\phi^{\alpha<}_\cmm(t)\ket 0 = \pi_{\alpha<}^\cmm(t)\ket 0 =
q^\mu_<(t)\ket 0 = p_\mu^<(t)\ket0 = 0.
\label{LER}
\ee
Then $DGRO(N,\oj)$ acts on the space of functionals
$C( q^\mu_>, p_\mu^>,\phi^{\alpha>}_\cmm,\pi_{\alpha>}^\cmm)$
of the remaining oscillators; this is the Fock module.
Define numbers $k_0(\rep)$, $k_1(\rep)$, $k_2(\rep)$ and $y_M$ by
\bes
\trrep T^\mu_\nu &=& k_0(\rep) \dlt^\mu_\nu, \nl
\trrep T^\mu_\nu T^\si_\tau &=&
 k_1(\rep) \dlt^\mu_\tau \dlt^\si_\nu
 + k_2(\rep) \dlt^\mu_\nu \dlt^\si_\tau, \\
\trM J_aJ_b &=& y_M \dlt_{ab}.
\eens
For an unconstrained tensor with $p$ upper and $q$ lower indices and
weight $\ka$, we have
\bes
\dim(\rep) = N^{p+q}, &\quad&
k_0(\rep)= -(p-q-\ka N) N^{p+q-1},
\label{krep}\\
k_1(\rep) = (p+q)N^{p+q-1}, &\quad&
k_2(\rep) = ((p-q-\ka N)^2 - p - q) N^{p+q-2}.
\eens
Note that if $\ka = (p-q)/N$, $\rep$ is an $sl(N)$ representation.
For the symmetric representations on $\ell$ lower indices, $S_\ell$,
and on $\ell$ upper indices, $S^\ell$, we have
\bes
\dim(S_\ell) = \dim(S^\ell) &=& {N-1+\ell\choose\ell}, \nl
k_0(S_\ell) = -k_0(S^\ell) &=& {N-1+\ell\choose\ell-1},
\nlb{kSl}
k_1(S_\ell) = k_1(S^\ell) &=& {N+\ell\choose\ell-1}, \nl
k_2(S_\ell) = k_2(S^\ell) &=& {N-1+\ell\choose\ell-2}.
\eens

The values of the abelian charges $c_1 - c_5$ (\ref{DGRO})
were calculated in \cite{Lar98}, Theorems 1 and 3, and in
\cite{Lar01}, Theorem 1:
\bes
c_1 &=& 1 - u\Np{} - x {N+p+1\choose N+2}, \nl
c_2 &=& -v\Np{} - 2w \Np{+1} - x {N+p\choose N+2}, \nl
c_3 &=& 1 + (1-2\la) ( w\Np{} + x\Np{+1}),
\label{cs}\\
c_4 &=& 2N - x(1-6\la+6\la^2)\Np{}, \nl
c_5 &=& y\Np{}.
\eens
where
\bes
u = \mp k_1(\rep)\, \dim\,M, &\qquad&
x = \mp \dim\,\rep\,\dim\,M, \nl
v = \mp k_2(\rep)\, \dim\,M, &\qquad&
y = \mp \dim\,\rep\,y_M,
\label{numdef}\\
w = \mp k_0(\rep)\, \dim\,M, &&
\eens
and the sign factor depends on the Grassmann parity of $\fa$; the upper
sign holds for bosons and the lower for fermions, respectively.
The $p$-independent contributions to $c_1$, $c_3$ and $c_4$ come from
the trajectory $q^\mu(t)$ itself.

\section{MCCQ: Manifestly Covariant Canonical Quantization}
\label{sec:MCCQ}

In the previous sections we constructed representations of gauge and
diffeomorphism algebras. These representations are of lowest-energy
type, i.e. there is a Hamiltonian whose eigenvalues are bounded from
below; this is the kind of representations relevant to quantum theory.
Now it is time to apply them to physics. To this end, we first
reformulate quantum physics in the history phase space, which is the
representation theory's natural habitat.

Consider a classical dynamical system with action $S$ and degrees of freedom
$\fa$. As is customary in the antifield literature, we use an abbreviated
notation where the index $\alpha$ stands for both discrete indices and
spacetime coordinates.
Dynamics is governed by the Euler-Lagrange (EL) equations,
\be
\Ea = \da S \equiv {\dlt S\/\dlt\fa} = 0.
\label{EL}
\ee
An important role is also played by the Hessian, i.e. the symmetric
second functional-derivative matrix
\be
\Kab = K_{\bt\al} = \db\Ea = {\dlt \Ea\/\dlt\fb}
= {\dlt^2 S\/\dlt\fa\dlt\fb}.
\label{Hess}
\ee
The Hessian is assumed non-singular, so it has an
inverse $\Mab$ satisfying
\be
K_{\bt\gm}M^{\gm\al} = M^{\al\gm}K_{\gm\bt} = \dlt^\al_\bt.
\ee
Introduce an antifield $\fsa$ for each EL equation (\ref{EL}), and
replace the space of $\phi$-histories $\QQ$ by the extended history
space $\QQ^*$, spanned by both $\phi$ and $\fs$.
In $\QQ^*$ we define the Koszul-Tate (KT) differential $\dlt$ by
\be
\dlt \fa = 0, \qquad
\dlt \fsa = \Ea.
\label{dlt0}
\ee
One checks that $\dlt$ is nilpotent, $\dlt^2 = 0$.
Define the antifield number $\afn \fa = 0$, $\afn \fsa = 1$. The KT
differential clearly has antifield number $\afn\dlt = -1$.

The space $C(\QQ^*)$ decomposes into subspaces $C^k(\QQ^*)$ of
fixed antifield number
\be
C(\QQ^*) = \sum_{k=0}^\infty C^k(\QQ^*)
\ee
The KT complex is
\be
0 \larroww \dlt C^0 \larroww \dlt C^1 \larroww \dlt
C^2 \larroww \dlt \ldots
\label{complex1}
\ee
The cohomology spaces are defined as usual by
$H_\cl^\bullet(\dlt) = \ker\dlt/\im\dlt$, i.e.
$H_\cl^k(\dlt) = (\ker\dlt)_k/(\im\dlt)_k$, where
the subscript $\cl$ indicates that we deal with a classical phase space.
It is easy to see that
\bes
(\ker \dlt)_0 &=& C(\QQ), \nle
(\im \dlt)_0 &=& C(\QQ)\Ea \equiv \N.
\eens
Thus $H_\cl^0(\dlt) = C(\QQ)/\N = C(\Sigma)$.
Since we assume that there are no non-trivial relations among the
$\Ea$, the higher cohomology groups vanish. This is a standard
result \cite{HT92}. The complex (\ref{complex1}) thus gives us
a resolution of the covariant phase space $C(\Sigma)$, which by
definition means that
$H_\cl^0(\dlt) = C(\Sigma)$, $H_\cl^k(\dlt) = 0$, for all $k>0$.

Alas, the antifield formalism is not suited for canonical quantization.
We can define an antibracket in
$\QQ^*$, but in order to do canonical quantization we need an honest
Poisson bracket. To this end, we introduce canonical momenta conjugate to
the history and its antifield, and obtain an even larger space $\PP^*$,
which may be thought of as the phase space corresponding to the
extended history space $\QQ^*$.

Introduce canonical momenta $\pa = \dd{\fa}$ and
$\psa = \dd{\fsa}$ for both the fields and antifields. The momenta
satisfy by definition the graded canonical commutation relations
($\fa$ is assumed bosonic),
\bes
[\pb,\fa] = \dlt^\al_\bt, &\qquad&
[\fa,\fb] = [\pa,\pb] = 0,
\nlb{ccr*}
\{\psb,\fsa\} = \dlt_\al^\bt, &\qquad&
\{\fsa,\fsb\} = \{\psa,\psb\} = 0,
\eens
where $\{\cdot,\cdot\}$ is the symmetric bracket.
Let $\PP$ be the phase space of histories with basis $(\fa,\pb)$,
and let $\PP^*$ be the extended phase space with basis
$(\fa,\pb,\fsa,\psb)$.

The definition of the KT differential extends to $\PP^*$ by
requiring that $\dlt F = [Q_{KT},F]$ for every $F \in C(\PP^*)$,
where the KT operator is
\be
Q_{KT} = \Ea \psa.
\ee
It acts on the various fields as
\bes
\dlt \fa &=& 0, \nl
\dlt \fsa &=& \Ea,
\nlb{dltfp}
\dlt \pa &=& -{\dlt \Eb\/\dlt\fa}\psb = - \Kab \psb, \nl
\dlt \psa &=& 0,
\eens
where $\Kab$ is the Hessian (\ref{Hess}).
We check that $\dlt$ is still nilpotent: $\dlt^2 = \{Q_{KT},Q_{KT}\} = 0$.

Like $C(\QQ)$, the function space $C(\PP^*)$
decomposes into subspaces of fixed antifield number,
$C(\PP^*) = \sum_{k=-\infty}^\infty C^k(\PP^*)$.
We can therefore define a KT complex in $C(\PP^*)$
\be
\ldots \larroww \dlt C^{-2} \larroww \dlt C^{-1} \larroww \dlt
C^0 \larroww \dlt C^1 \larroww \dlt
C^2 \larroww \dlt \ldots
\label{complex2}
\ee
Because the Hessian (\ref{Hess}) is non-singular by assumption
with inverse $\Mab$, we can invert the relation
$\dlt \pa = -\Kab\psb$
and get
\be
\psa = -\Mab\dlt\pb = \dlt(-\Mab\pb),
\ee
since $\Mab$ depends on $\phi$ alone.

Let us now compute the cohomology.
Any function which contains $\pa$ is not closed, so
$\ker\dlt = C(\phi,\fs,\ps)$.
Moreover, $\im\dlt$ is generated by the two ideals
$\Ea$ and $\psa$. The momenta $\pa$ and $\psa$ thus vanish
in cohomology, and the part with zero antifield number is thus still
$H_\cl^0(\dlt) = C(\QQ)/\N = C(\Sigma)$.
The higher cohomology groups $H_\cl^k(\dlt) = 0$ by the same argument
as above. Hence the complex (\ref{complex2}) yields a different
resolution of the function space $C(\Sigma)$.

It is important that the spaces $C^k$ in (\ref{complex2}) are phase
spaces, equipped with the Poisson bracket (\ref{ccr*}). Unlike the
resolution (\ref{complex1}), the new resolution (\ref{complex2}) therefore
allows us to do canonical quantization: replace Poisson brackets by
commutators and represent the graded Heisenberg algebra (\ref{ccr*}) on
a Hilbert space. However, the Heisenberg algebra can be represented on
different Hilbert spaces; there is no Stone-von Neumann theorem in
infinite dimension. To pick the correct one, we must impose the
physical condition that there is an energy which is bounded on below.

To define the Hamiltonian, we must single out a privileged variable
$t$ among the $\al$'s, and declare it to be time.
Thus replace $\al = (i,t)$, so e.g. $\fa = \phi^i(t)$,
$\Ea = \EE_i(t)$, etc.
This step means of course that we sacrifice covariance.
The Hamiltonian reads
\be
H = -i\int dt\ \dot\phi^i(t)\pi_i(t) + \dot\fsi(t)\ps^i(t).
\label{Hamc}
\ee
It satisfies
\bes
[H, \phi^i(t)] = -i\dot\phi^i(t), &\qquad&
[H, \pi_i(t)] = -i\dot\pi_i(t), \nle
{[}H, \fsi(t)] = -i\dot\fsi(t), &\qquad&
[H, \ps^i(t)] = -i\dot\ps^i(t).
\eens

Expand all fields in a Fourier series with respect to time, e.g,
\be
\phi^i(t) = \intdm \phi^i(m) \e^{imt}.
\ee
The Fourier modes $\pi_i(m)$, $\fsi(m)$ and $\ps^i(m)$ are defined
analogously.
The Hamiltonian acts on the Fourier modes as
\bes
[H, \phi^i(m)] = m\phi^i(m), &\qquad&
[H, \pi_i(m)] = m\pi_i(m), \nle
{[}H, \fsi(m)] = m\fsi(m), &\qquad&
[H, \ps^i(m)] = m\ps^i(m).
\eens

Now quantize. In the spirit of BRST quantization, our strategy is to
quantize first and impose dyna\-mics afterwards.
In the extended history phase space $\PP^*$, we
define a Fock vacuum $\ket 0$ which is annihilated by all negative
frequency modes, i.e.
\be
\phi^i(-m)\ket 0 = \pi_i(-m)\ket 0 = \fsi(-m)\ket 0
= \ps^i(-m)\ket 0 = 0,
\ee
for all $-m < 0$. We must also decide which of the zero modes that
annihilate the vacuum, but the decision is not important unless
zero-momentum modes will survive in cohomology, and even then it will
not affect the eigenvalues of the Hamiltonian.

The Hamiltonian (\ref{Hamc}) does not act in a well-defined manner,
because it assigns an infinite energy to the Fock vacuum. To correct
for that, we replace the Hamiltonian by
\be
H = -i\int dt\ \no{\dot\phi^i(t)\pi_i(t)} + \no{\dot\fsi(t)\ps^i(t)},
\label{Hamq}
\ee
where normal ordering $\no{\cdot}$ moves negative frequency modes to
the right and positive frequency modes to the left. The vacuum has
zero energy as measured by the normal-ordered Hamiltonian, $H\ket0 = 0$.
The Hilbert space can be identified with
\be
\HH(\PP^*) = C(\phi^i(m>0), \pi_i(m>0), \fsi(m>0), \ps^i(m>0)).
\ee
The energy of a state in $\HH(\PP^*)$ follows from
\be
H \phi^{i_1}(m_1)...\ps^{i_n}(m_n)\ket 0
= (m_1 + ... + m_n)  \phi^{i_1}(m_1)...\ps^{i_n}(m_n)\ket 0.
\label{Hnoncov}
\ee

It is important that the KT operator
\be
Q_{KT} = \Ea\psa = \int dt\ \EE_i(t) \ps^i(t) = \intdm \EE_i(m)\ps^i(-m)
\ee
is already normal ordered, because $\Ea$ and $\psa$ commute.
This means that $Q_{KT}^2 = 0$ also quantum mechanically; there are no
anomalies. Moreover, $Q_{KT}$ still commutes with the Hamiltonian,
$[Q_{KT}, H] = 0$, and this property is not destroyed by normal
ordering.
Hence the Hilbert space $\HH(\PP^*)$ has also a well-defined
decomposition into subspaces of definite antifield number,
\be
\HH(\PP^*) = ... + \HH^{-2} + \HH^{-1} + \HH^0 + \HH^1 + \HH^2 + ...
\ee
There is a KT complex in $\HH(\PP^*)$
\be
\ldots \larroww \dlt \HH^{-2} \larroww \dlt \HH^{-1} \larroww \dlt
\HH^0 \larroww \dlt \HH^1 \larroww \dlt
\HH^2 \larroww \dlt \ldots
\label{complexq}
\ee
The physical Hilbert space is identified with
$\HH(\Sigma) = \ab \HH_\qm^0(Q_{KT}) =
\break (\ker Q_{KT})_0/(\im Q_{KT})_0$.
The action of the Hamiltonian on the physical Hilbert space is still
given by (\ref{Hnoncov}), restricted to $\HH(\Sigma)\subset\HH(\PP^*)$,
and that coincides with the conventional action of the Hamiltonian.

Hence we have quantized the theory given by the EL equation (\ref{EL})
by first quantizing the space of phase space histories $\PP^*$, and
then imposing dyna\-mics through KT cohomology.

\section{Scalar field I: non-covariant quantization}
\label{sec:Scalar1}

The action, Euler-Lagrange equations, and Hessian read
\be
S &=& \half \int \dNx\  (\dmu \phi(x)\d^\mu \phi(x) - \om^2 \phi^2(x)), \nl
\EE(x) &\equiv& -{\dlt S\/\dlt \phi(x)}
= \Box \phi(x) + \omega^2 \phi(x) = 0,
\nlb{Sscalar}
K(x,x') &\equiv& -{\dlt^2 S\/\dlt \phi(x)\dlt \phi(x')}
= \Box \delta(x-x') + \omega^2 \delta(x-x'),
\eens
where $\Box = \dmu \d^\mu$.

Introduce antifields $\fs(x)$ and canonical momenta
$\pi(x) = \dd{\phi(x)}$ and  $\ps(x) = \dd{\fs(x)}$.
The non-zero brackets are
\be
[\pi(x),\phi(x')] = \{\ps(x),\fs(x')\} = \dlt(x-x').
\ee
The KT differential reads
\be
Q_{KT} = \int \dNx\  (\Box \phi(x) + \omega^2 \phi(x))\ps(x).
\ee
$Q_{KT}$ acts as $\dlt F = [Q_{KT},F]$, where
\bes
\dlt \phi(x) &=& 0, \nl
\dlt \fs(x) &=& \Box \phi(x) + \omega^2 \phi(x), \nle
\dlt \pi(x) &=& -(\Box \ps(x) + \omega^2 \ps(x)), \nl
\dlt \ps(x) &=& 0.
\eens

Now we do a Fourier transformation. The extended phase space $\PP^*$
is spanned by modes $\phi(k)$, $\fs(k)$, $\pi(k)$ and $\ps(k)$, and
the EL equation becomes
\be
\EE(k) = -(k^2 - \omega^2)\phi(k) = 0.
\ee
The non-zero brackets are
\be
[\pi(k), \phi(k')] = \{\ps(k), \fs(k')\} = \dlt(k+k').
\ee
The KT differential is
\be
Q_{KT} = \int \dNk\ (k^2 - \omega^2) \phi(k) \ps(-k).
\ee
$Q_{KT}$ acts as $\dlt F = [Q_{KT},F]$, where
\bes
\dlt \phi(k) &=& 0, \nl
\dlt \fs(k) &=& (k^2 - \omega^2)\phi(k),
\nlb{Qqp}
\dlt \pi(k) &=& -(k^2 - \omega^2) \ps(k), \nl
\dlt \ps(k) &=& 0.
\eens

The cohomology is computed as follows.
Since the equations (\ref{Qqp}) decouple, we can consider each value
of $k$ separately. First assume that $k^2 \neq \omega^2$.
$\phi(k)$ and $\ps(k)$ are closed for all $k$, but $\fs(k)$ and $\pi(k)$
are not closed since $\dlt \fs(k) \neq 0$, etc.
We can invert the second and third equations to read
\bes
\phi(k) &=& {1\/k^2 - \omega^2} \dlt \fs(k), \nle
\ps(k) &=& -{1\/k^2 - \omega^2} \dlt \pi(k).
\eens
Hence $\phi(k)$ and $\ps(k)$ lie in the image of $\dlt$,
and the cohomology vanishes completely: only $\phi(k)$ and $\ps(k)$ lie in
the kernel, but they also lie in the image.

Now turn to the case $k^2 = \omega^2$, say $k=(\om,0,0,0)$. Clearly,
$\dlt \phi(k) = \dlt \pi(k) = \dlt \fs(k) = \dlt \ps(k) = 0$, so
all four variables lie in the kernel but not in the image.
Thus the cohomology spaces are too big;
the classical cohomology spaces can be identified with
$H_\cl^\bullet(\dlt) = C(\phi(k), \pi(k), \fs(k),\ps(k))$.
The zeroth cohomology space consists of such functions
with total antifield number zero, i.e.
$H_\cl^0(\dlt) = C(\phi(k), \pi(k), (\fs(k)\ps(k')))$.
In \cite{Lar04} it was proposed that this problem could be handled by
adding a small perturbation to make the Hessian non-singular, so the
momenta can be killed in cohomology. This unwanted cohomology is
an embarassment, especially since it reappears in Maxwell theory,
but we have nothing more to say about it.

To quantize the theory we must specify a Hamiltonian. Let it be
\bes
H &=& -i\int \dNx\  (\d_0 \phi(x)\pi(x) + \d_0 \fs(x)\ps(x)) \nle
  &=& \int \dNk\ k_0 (\phi(k) \pi(-k) + \fs(k) \ps(-k)).
\eens
Note that at this stage we break Poincar\'e invariance,
since the Hamiltonian treats the $x^0$ coordinate differently from
the other $x^\mu$.
Quantize by introducing a Fock vacuum $\ket 0$ satisfying
\be
\phi(k)\ket 0 = \pi(k)\ket 0 = \fs(k)\ket 0 = \ps(k)\ket 0 = 0,
\ee
for all $k$ such that $k_0 < 0$.
After adding a small
perturbation to make the Hessian invertible, $\pi(k)$ and $\ps(k)$
vanish in cohomology, as do the off-shell components of $\phi(k)$
and $\fs(k)$. The classical cohomology
$H_\cl^\bullet(Q_{KT}) = C(\phi(k; k^2 = \om^2), \fs(k; k^2 = \om^2))$
consists of functions of the on-shell components of $\phi$ and $\fs$, and
$H_\cl^0(Q_{KT}) = C(\phi(k; k^2 = \om^2))$ is the classical phase space.
The quantization step eliminates the components $\phi(k)$ with $k_0 < 0$,
which leaves us with the physical Hilbert space
$\HH = H_\qm^0(Q_{KT}) = C(\phi(k; k^2 = \om^2\ \hbox{and}\ k_0 > 0))$.
A basis for $\HH$ consists of multi-quanta states
\be
\ket{k, k', ..., k^{(n)}} = \phi(k)\phi(k')...\phi(k^{(n)})\ket 0
\ee
with energy $H = k + k' + ... + k^{(n)}$.

\section{MCCQ: Jets and covariant quantization}
\label{sec:MCCQ-jets}

A covariant definition of the phase space was given in the Section
\ref{sec:MCCQ}, but the Hamiltonian and thus the quantum Hilbert space
broke covariance, due to the selection of a privileged time coordinate.
In this section we correct this defect.

The compact notation is not very useful
here, because the notion of covariance does not make sense unless
some indices are identified with spacetime coordinates. So we assume
that we have some fields $\fa(x)$, where $x = (x^\mu) \in \RR^N$ is
the spacetime coordinate. The EL equations read
\be
\Ea(x) \equiv {\dlt S\/\dlt \fa(x)} = 0.
\ee
We also need the Hessian
\be
\Kab(x,x') = K_{\bt\al}(x',x)
= {\dlt \Ea(x)\/\dlt\fb(x')}
= {\dlt^2 S\/\dlt\fa(x)\dlt\fb(x')}.
\label{Hess2}
\ee
which we assume is non-singular.

Now let all fields depend on an additional parameter $t$. It will
eventually be identified with time, but so far it is completely
unrelated to the $x^\mu$.
Upon the substitution $\fa(x) \to \fa(x,t)$,
the EL equations are replaced by
\be
\Ea(x,t) = 0.
\label{EL2}
\ee
The Hessian (\ref{Hess2}) becomes
\be
\Kab(x,t,x',t') = K_{\bt\al}(x',t',x,t)
= {\dlt \Ea(x,t)\/\dlt\fb(x',t')},
\ee
which has the inverse
$\Mab(x,t,x',t')$ satisfying
\[
\int\dNxb\int dt''\ K_{\bt\gm}(x,t,x'',t'')M^{\gm\al}(x'',t'',x',t')
= \dlt^\al_\bt \dlt(x-x') \dlt(t-t').
\]

To remove the condition (\ref{EL2}) in cohomology we introduce
antifields $\fsa(x,t)$.
But the fields in the physical phase space do not depend on the parameter
$t$, which gives rise to the extra condition
\be
\d_t\fa(x,t) \equiv {\d\fa(x,t)\/\d t} = 0.
\ee
We can implement this condition by introducing new antifields $\wfa(x,t)$.
However, the identities $\d_t\Ea(x,t) \equiv 0$ give rise to unwanted
cohomology. To kill this condition, we must introduce yet another
antifield $\wfsa(x,t)$.
The KT differential $\dlt$ is defined by
\bes
\dlt \fa(x,t) &=& 0, \nl
\dlt \fsa(x,t) &=& \Ea(x,t),
\nlb{dltsi}
\dlt \wfa(x,t) &=& \d_t\fa(x,t), \nl
\dlt \wfsa(x,t) &=& \d_t\fsa(x,t)
- \int\dNxp \int dt'\ \Kab(x,t,x',t') \wfb(x',t').
\eens
The zeroth cohomology group $H_\cl^0(\dlt)$ equals $C(\phi)$, modulo
the ideals generated by $\Ea(x,t)$ and $\d_t\fa(x,t)$.
Moreover, the wouldbe cohomology related to the identity
\be
\dlt \Big\{\d_t\fsa(x,t) -
\int\dNxp \int dt'\ {\dlt \Ea(x,t)\/\dlt\fb(x',t')} \wfb(x',t') \Big\}
\equiv 0
\ee
is killed because the expression equals $\dlt\wfsa(x,t)$.

Introduce canonical momenta for all fields and antifields:
$\pa(x,t) = \dd{\fa(x,t)}$, $\psa(x,t) = \dd{\fsa(x,t)}$,
$\wpa(x,t) = \dd{\wfa(x,t)}$, and $\wpsa(x,t) = \dd{\wfsa(x,t)}$.
The KT operator takes the explicit form
\be
Q_{KT} &=& \int \dNx \int dt\ \Big\{ \Ea(x,t))\psa(x,t) + \d_t\fa(x,t)\wpa(x,t)
\\
&&+( \d_t\fsa(x,t)
- \int\dNxp \int dt'\ \Kab(x,t,x',t') \wfb(t'))\wpsa(x,t) \Big\}.
\eens
From this we can read off the action of $\dlt$ on the momenta.
As in the previous section, the zeroth cohomology group consists of
functions $\fa(x,t)$ which satisfy $\Ea(x,t) = 0$ and $\d_t\fa(x,t)=0$.
Hence $H_\cl^0(\dlt) = C(\Sigma)$, as desired.

At this point, we must define a Hamiltonian. The candidate
\bes
H_0 &=& -i\int dt\ \Big\{\d_t\fa(x,t)\pa(x,t) + \d_t\fsa(x,t)\psa(x,t)
\nlb{Hamconstr}
&&+ \d_t\wfa(x,t)\wpa(x,t) + \d_t\wfsa(x,t)\wpsa(x,t) \Big\}
\eens
might seem natural, but it is not acceptable. The action of the
Hamiltonian is KT exact, e.g.
\be
[H_0, \fa(x,t)] = \d_t\fa(x,t) = \dlt\wfa(x,t),
\ee
and thus $H_0\approx0$. $H_0$ is not a genuine Hamiltonian, but
rather a Hamiltonian constraint $H_0\approx0$, familiar from canonical
quantization of general relativity.

However, we can construct a well-defined and physical Hamiltonian with
some extra work. The crucial idea is to introduce the observer's
trajectory $q^\mu(t) \in \RR^N$, and then expand all fields in a Taylor
series around this trajectory as in (\ref{Taylor}).
Expand also the Euler-Lagrange equations and the antifields in
a similar Taylor series, e.g.
$\Ea(x,t) = \sum_{\mm} {1\/\mm!} \Eam(t)(x-q(t))^\mm$.
Such relations define the jets $\Eam(t)$, $\fsam(t)$, $\wfam(t)$
and $\wfsam(t)$. Jets of antifields will sometimes be called antijets.

The equation of motion and the time-independence condition translate
into
\bes
\Eam(t) &=& 0, \nle
D_t\fam(t) &\equiv& \dotfam(t)
- \sum_\mu \dot q^\mu(t)\fa_{\cmm+\mu}(t) = 0.
\eens
The KT differential $\dlt$ which implements these conditions is
\bes
\dlt \fam(t) &=& 0, \nl
\dlt \fsam(t) &=& \Eam(t),
\nle
\dlt \wfam(t) &=& D_t\fam(t), \nl
\dlt \wfsam(t) &=& D_t\fsam(t)
 - \sum_\nn \int dt'\ K^\nn_{\mm;\al\bt}(t,t') \wfbn(t').
\eens
The cohomology group $H_\cl^0(\dlt)$ consists of linear combinations
of jets $\fam(t)$ satisfying $\Eam(t)=0$ and $D_t\fam(t)=0$.

The Taylor expansion requires that we introduce the observer's
trajectory as a physical field, but what equation
of motion does it obey? The obvious answer is the geodesic equation,
which we compactly write as $\GG_\mu(t) = 0$. The geodesic operator
$\GG_\mu(t)$ is a function of the metric $g_{\mu\nu}(q(t),t)$
and its derivatives on the curve $q^\mu(t)$. To eliminate this ideal
in cohomology we introduce the trajectory antifield $\qsmu(t)$, and
extend the KT differential to it:
\bes
\dlt  q^\mu(t) &=& 0,
\nlb{geo}
\dlt \qsmu(t) &=& \GG_\mu(t).
\eens
For models defined over Minkowski spacetime, the geodesic equation
simply becomes $\ddot q^\mu(t) = 0$, and the KT differential reads
\be
\dlt \qsmu(t) = \eta_{\mu\nu} q^\nu(t).
\ee
$H_\cl^0(\dlt)$ only contains trajectories which are straight lines,
\be
 q^\mu(t) = u^\mu t + a^\mu,
\label{line}
\ee
where $u^\mu$ and $a^\mu$ are constant vectors. We may also require
that $u^\mu$ has unit length, $u_\mu u^\mu = 1$. This condition fixes
the scale of the parameter $t$ in terms of the Minkowski metric, so
we may regard it as proper time rather than as an arbitrary parameter.

Now introduce the canonical momenta $\pam(t) = \dd{\fam(t)}$,
$\psam(t) = \dd{\fsam(t)}$, $\wpam(t) = \dd{\wfam(t)}$,
$\wpsam(t) = \dd{\wfsam(t)}$ for the jets and antijets (jet and antijet
momenta), and momenta $p_\mu(t) = \dd{q^\mu(t)}$ and
$\psmu(t) = \dd{\qsmu(t)}$ for the observer's trajectory
and its antifield. We can now define a
genuine Hamiltonian $H$, which translates the fields relative to the
observer or vice versa. Since the formulas are shortest when $H$ acts
on the trajectory but not on the jets, we make that choice, and define
\be
H = i\int dt\ (\dot q^\mu(t) p_\mu(t) + \dot\qsmu(t)\psmu(t)).
\label{H2}
\ee
Note the sign; moving the fields forward in $t$ is equivalent to moving
the observer backwards.
{F}rom (\ref{Taylor}) we get the energy of the fields:
\be
[H, \fa(x,t)] = -i\dot q^\mu(t)\dmu\fa(x,t).
\label{Hcl}
\ee
This a crucial result, because it allows us to define a
genuine energy operator in a covariant way.
In Minkowski space, the trajectory is a straight line (\ref{line}),
and $\dot q^\mu(t) = u^\mu$. If we take $u^\mu$ to be the constant
four-vector $u^\mu = (1,0,0,0)$, then (\ref{Hcl}) reduces to
\be
[H, \fa(x,t)] = -i {\d\/\d x^0}\fa(x,t).
\ee
Equation (\ref{H2}) is thus a genuine covariant generalization of
the energy operator.

Now we quantize the theory.
Since all operators depend on the parameter $t$, we can define
the Fourier components as in (\ref{Fourier}).
The the Fock vacuum $\ket 0$ is defined to be annihilated by
all negative frequency modes, $\fam(-m)$, $q^\mu(-m)$, etc. with $m<0$.
The normal-ordered form of the Hamiltonian (\ref{H2}) reads, in
Fourier space,
\be
H = -\intdm m( \no{ q^\mu(m) p_\mu(-m)} + \no{\qsmu(m) p_\mu(-m)} ),
\label{Hq}
\ee
where double dots indicate normal ordering with respect to frequency.
This ensures that $H\ket 0 = 0$.
The classical phase space $H_\cl^0(\dlt)$ is thus the the space of
fields $\fa(x)$ which solve $\Ea(x)=0$,
and trajectories $q^\mu(t) = u^\mu t+a^\mu$, where $u^2 = 1$.
After quantization, the fields and trajectories become operators which
act on the physical Hilbert space $\HH = H_\qm^0(Q_{KT})$, which is the space
of functions of the positive-energy modes of the classical phase
space variables.

This construction differs technically from conventional canonical
quantization, but there is also a physical difference. Consider the
state $\ket{\fa(x)} = \fa(x)\ket 0$ which excites one $\phi$ quantum
from the vacuum. The Hamiltonian yields
\bes
H \ket{\fa(x)} &=& -i\dot q^\mu(t)\dmu\fa(x)\ket 0 \nl
&=& -i\ket{\dot q^\mu(t)\dmu\fa(x)} \\
&=& -i\ket{u^\mu\dmu\fa(x)}.
\eens
If $u^\mu$ were a classical variable, the state $\ket{\fa(x)}$ would
be a superposition of energy eigenstates:
\be
H \ket{\fa(x)} = -iu^\mu\dmu\ket{\fa(x)}.
\ee
In particular, let $u^\mu = (1,0,0,0)$ be a unit vector in the $x^0$
direction and $\fa(x) = \exp(ik\cdot x)$ be a plane wave. We then
define the state $\ket{0;u,a}$ by
\be
 q^\mu(t)\ket{0;u,a} = (u^\mu t + a^\mu)\ket{0;u,a}.
\label{ua}
\ee
Now write $\ket{k; u,a} =\exp(ik\cdot x)\ket{0;u,a}$ for
the single-quantum energy eigenstate.
\be
H \ket{k; u,a} = k_\mu u^\mu\ket{k;u,a},
\ee
so the eigenvalue of the Hamiltonian is $k_\mu u^\mu = k_0$, as
expected. Moreover, the lowest-energy condition ensures
that only quanta with positive energy will be excited; if
$k_\mu u^\mu < 0$ then $\ket{k; u,a} = 0$.

However, the present analysis shows that it is in principle wrong to
consider $u^\mu$ and $a^\mu$ as classical variables. The definition
(\ref{ua}) means that the reference state $\ket{0;u,a}$ is a very
complicated, mixed, macroscopic state where the observer moves along a
well-defined, classical trajectory. This is of course an excellent
approximation in practice, but in principle wrong.

\section{Scalar field II: covariant quantization}
\label{sec:Scalar2}

Following the prescription in Section \ref{sec:MCCQ-jets}, we make the
replacement $\phi(x) \to \phi(x,t)$, where $t\in\RR$ is a parameter.
The EL equation (\ref{Sscalar}) becomes
\be
\EE(x,t) \equiv \Box \phi(x,t) + \omega^2 \phi(x,t) = 0.
\ee
To remove this condition in cohomology we introduce antifields $\fs(x,t)$.
But there is an extra condition
\be
\d_t\phi(x,t) \equiv {\d\phi(x,t)\/\d t} = 0.
\ee
We can implement this condition by introducing new antifields $\wf(x,t)$.
However, the identities $\d_t\EE(x,t) \equiv 0$ give rise to unwanted
cohomology. To kill this condition, we must introduce a second-order
antifield $\wfs(x,t)$.
After passage to jet space, the equation of motion and the
time-independence condition translate into
\bes
\sum_\mu \phi_{\cmm+2\mu}(t) + \omega^2 \fm(t) &=& 0, \nle
D_t\fm(t) \equiv \dotfm(t) - \sum_\mu \dot q^\mu(t)\phi_{\cmm+\mu}(t) &=& 0.
\eens
We introduce anti-jets $\fsm(t)$, $\wfm(t)$ and $\wfsm(t)$ and the
KT differential $\dlt$ to implement these conditions:
\bes
\dlt \fm(t) &=& 0, \nl
\dlt \fsm(t) &=& \sum_\mu \phi_{\cmm+2\mu}(t) + \omega^2 \fm(t),
\nle
\dlt \wfm(t) &=& D_t\fm(t), \nl
\dlt \wfsm(t) &=& D_t\fs(t)
 - (\sum_\mu \wf_{\cmm+2\mu}(t) + \omega^2 \wfm(t)).
\eens
The classical cohomology group $H_\cl^0(\dlt)$ is spanned by of linear
combinations of jets satisfying
\be
\fm(t) = \e^{ik\cdot q(t)} (ik)^\mm
\label{fmt}
\ee
where $k^2 = \om^2$, $k\cdot q = k_\mu q^\mu$ and the power
$k^\mm$ is defined in analogy with (\ref{power}).
It is hardly surprising that the Taylor series can be summed, giving
\bes
\phi(x,t) &=&
\e^{ik\cdot q(t)} \sum_{\mm} {1\/\mm!} (ik)^\mm(x-q(t))^\mm \nl
&=& \e^{ik\cdot q(t)} \e^{ik\cdot(x-q(t))}
\label{phixt}\\
&=& \eikx.
\eens

The physical Hamiltonian $H$ is defined as in Equation (\ref{Hq}). The
classical phase space $H_\cl^0(\dlt)$ is thus the the space of plane waves
$\eikx$, cf (\ref{phixt}),  and trajectories $q^\mu(t) = u^\mu t+a^\mu$.
The energy is given by
\bes
[H, \eikx] &=& k_\mu \dot q^\mu(t)\eikx = k_\mu u^\mu \eikx, \nle
[H,  q^\mu(t)] &=& i\dot q^\mu(t).
\eens
This is a covariant description of phase space, because the energy
$k_\mu u^\mu$ is Poincar\'e invariant.

We now quantize the theory before imposing dyna\-mics. To this end, we
introduce the canonical momenta $\pim(t)$, $\psm(t)$, $\wpm(t)$,
$\wpsm(t)$ for the jets and antijets, and $p_\mu(t)$ and $\psmu(t)$
for the observer's trajectory and its antifield.
Since the jets also depend on the parameter $t$, we can define
their Fourier components as in (\ref{Fourier}).
The the Fock vacuum $\ket 0$ is defined to be annihilated by
the negative frequency modes of the jets and antijets,
and the quantum Hamiltonian is still defined by (\ref{Hq}),
where double dots indicate normal ordering with respect to frequency,
ensuring that $H\ket 0 = 0$.

The rest proceeds as in the end of Section \ref{sec:MCCQ-jets}.
We can consider the one-quantum state with momentum $k$ over the
true Fock vacuum, $\ket k = \exp(ik\cdot x)\ket 0$. This state is
not an energy eigenstate, because the Hamiltonian excites a quantum
of the observers trajectory: $H\ket k = k_\mu u^\mu\ket k$.
We may think of the observer's trajectory as a classical
variable and introduce the macroscopic reference state
$\ket{0;u,a}$, on which
$q^\mu(t)\ket{0;u,a} = (u^\mu t + a^\mu)\ket{0;u,a}$.
We can then consider a state
$\ket{k; u,a} = \exp(ik\cdot x)\ket{k; u,a}$ with one quantum over
the reference state. The Hamiltonian gives
$H \ket{k; u,a} = k_\mu u^\mu\ket{k;u,a}$.
In particular, if $u^\mu = (1,0,0,0)$, then
the eigenvalue of the Hamiltonian is $k_\mu u^\mu = k_0$, as
expected. Moreover, the lowest-energy condition (\ref{LER}) ensures
that only quanta with positive energy will be excited; if
$k_\mu u^\mu < 0$ then $\ket{k;u,a} = 0$.

\section{MCCQ: Gauge symmetries}
\label{sec:MCCQ-gauge}

In the previous sections MCCQ was applied to the free scalar field.
However, it is mainly useful for theories with gauge symmetries, due to
its connection with the representation theory of gauge algebras
developed earlier. We now come to this case, and assume that there
are some relations between the EL equations (\ref{EL}). In other words,
let there be identities of the form
\be
r^\al_a\Ea \equiv 0,
\label{Rident}
\ee
where the $r^\al_a$ are some functionals of $\fa$.
The zeroth cohomology group  $H_\cl^0(\dlt) = C(\QQ)/\N = C(\Sigma)$
is not changed, but the higher cohomology groups no longer vanish, since
$\dlt(r^\al_a\fsa) = r^\al_a\Ea \equiv 0$. The standard method to kill
this unwanted cohomology is to introduce a bosonic second-order
antifield $\za$, so that $r^\al_a\fsa = \dlt\za$ is KT exact.
The differential (\ref{dlt0}) is thus modified to read
\bes
\dlt \fa &=& 0, \nl
\dlt \fsa &=& \Ea,
\label{dlt1}\\
\dlt \za &=& r^\al_a\fsa.
\eens
By introducing canonical momenta $\chi^a = \dd{\za}$ for the second-order
antifields, we can write the KT differential as a bracket, $\dlt F =
[Q_{KT},F]$, where the full KT operator is
\be
Q_{KT} =  \Ea\psa + r^\al_a\fsa\chi^a.
\label{KT}
\ee
$Q_{KT}$ is an operator in the extended phase space $\PP^*$
with basis $(\fa,\pb,\ab\fsa,\psb,\ab\za,\chi^b)$, and
$\{Q_{KT}, Q_{KT}\} = 0$.

The identity (\ref{Rident})
implies that $J_a = r^\al_a\pa$ generate a Lie algebra under the Poisson
bracket. Namely, all $J_a$'s preserve the action, because
\be
[J_a, S] = r^\al_a[\pa,S] = r^\al_a\Ea \equiv 0,
\ee
and the bracket of two operators which preserve some structure also
preserves the same structure.
We will only consider the case that the $J_a$'s generate a proper Lie
algebra $\oj$ as in (\ref{oj}).
The formalism extends without too much extra work to the more general
case of structure functions $f_{ab}{}^c(\phi)$, but we will not need
this complication here.
It follows that the functions $r^\al_a$ satisfy the identity
\be
\db r^\al_b r^\bt_a - \db r^\al_a r^\bt_b = f_{ab}{}^c r^\al_c.
\label{rident}
\ee
The Lie algebra $\oj$ also acts on the antifields:
\bes
[J_a, \fa] &=& r^\al_a, \nl
{[}J_a, \fsa] &=& -\da r^\bt_a \fsb
\label{Jf}\\
{[}J_a, \zb] &=& f_{ab}{}^c\zc.
\eens
In particular, it follows that $\fsa$ carries a $\oj$ representation
because it transforms in the same way as $\pa$ does.

Classically, it is always possible to reduce the phase space further,
by identifying points on $\oj$ orbits. To implement this additional
reduction, we introduce ghosts $c^a$ with anti-field number
$\afn c^a = -1$, and ghost momenta $b_a$ satisfying
$\{b_a, c^b\} = \dlt^b_a$.
The Lie algebra $\oj$ acts on the ghosts as
$[J_a, c^b] = -f_{ac}{}^b c^c$.
The full extended phase space, still denoted by $\PP^*$, is spanned by
$(\fa,\pb,\ab \fsa,\psb,\ab \za,\chi^b,\ab c^a,b_b)$.
The generators of $\oj$ are thus identified with the following vector
fields in $\PP^*$:
\bes
J_a &=& r^\al_a\pa -\da r^\bt_a\fsb\psa + f_{ab}{}^c\zc\chi^b
- f_{ab}{}^c c^b b_c
\nlb{Ja}
&=& J^{field}_a + J^{ghost}_a,
\eens
where $J^{ghost}_a = -f_{ab}{}^c c^b b_c$ and $J^{field}_a$ is the rest.

Now define the longitudinal derivative $d$ by
\bes
d c^a &=& -\half f_{bc}{}^a c^b c^c, \nl
d \fa &=& r^\al_a c^a, \nle
d \fsa &=& \da r^\bt_a \fsb c^a, \nl
d \za &=& - f_{ab}{}^c \zc c^b.
\eens
The longitudinal derivative can be written as $dF = [Q_{Long},F]$
for every $F\in C(\QQ^*)$, where
\be
Q_{Long} = J^{field}_a c^a - \half f_{ab}{}^c c^a c^b b_c
= J^{field}_a c^a + \half J^{ghost}_a c^a.
\label{QLong1}
\ee
We note that $Q_{Long}$ can be considered as smeared gauge generators,
$\J_X = X^a J_a$, where the smearing function $X^a$ is the fermonic
ghost $c^a$:
\be
Q_{Long} = \J^{field}_c + \half \J^{ghost}_c.
\label{smear}
\ee

One verifies that $d^2 = 0$ when acting on the fields and antifields by
means of the identify (\ref{rident}) and the Jacobi identities for $\oj$.
Moreover, it is straightforward to show that $d$ anticommutes with the
KT differential, $d\dlt = -\dlt d$; the proof is again done by checking
the action on the fields.
Hence we may define the nilpotent {\em BRST derivative} $s = \dlt+d$,
\bes
s c^a &=& -\half f_{bc}{}^a c^b c^c, \nl
s \fa &=& r^\al_a c^a,
\nlb{sfields}
s \fsa &=& \Ea + \da r^\bt_a \fsb c^a, \nl
s \za &=& r^\al_a\fsa - f_{ab}{}^c \zc c^b.
\eens
Nilpotency immediately follows because
$s^2 = \dlt^2 + \dlt d + d\dlt + d^2 = 0$.
The BRST operator can be written in the form $sF = [Q_{BRST},F]$ with
\bes
Q_{BRST} &=& Q_{KT} + Q_{Long} \nl
&=&  \Ea\psa + r^\al_a\fsa\chi^a + J^{field}_a c^a
+ \half J^{ghost}_a c^a \nle
&=&  -\half f_{ab}{}^c c^a c^b b_c + r^\al_a c^a \pa
+ (\Ea + \da r^\bt_a \fsb c^a) \psa \nl
&&+ (r^\al_a\fsa - f_{ab}{}^c \zc c^b) \chi^a.
\eens

In non-covariant quantization, we single out a privileged variable $t$
among the $\al$'s, and declare it to be time.
In the absense of gauge symmetries, the BRST operator reduces to the
KT operator (\ref{KT}), which is already normal ordered and hence
nilpotent on the quantum level. The question is whether the full BRST
operator also has this property.
The dangerous part is the longitudinal operator
\be
Q_{Long} = \int dt\ \Big\{ \no{J^{field}_a(t)} c^a(t)
+ \half \no{ J^{ghost}_a(t) c^a(t) } \Big\}.
\label{QLong2}
\ee
which ceases to be nilpotent unless the
normal-ordered gauge generators
$J_a(t) = \no{J^{field}_a(t)} + \no{ J^{ghost}_a(t)}$
generate the algebra (\ref{oj}) without additional quantum corrections.
If such an extension arises, the BRST operator ceases to be nilpotent.
However, the situation is even worse. Not only do quantum effects
generically ruin
nilpotency of the BRST operator, but they make the gauge generators ill
defined. However, it is possible to regularize the theory formulated in
terms of Taylor data, in such a way that the full gauge symmetry of the
original model is preserved, and the regularized gauge generators are
well-defined operators. The price to pay is the appearance of an anomaly.

The next step in Section \ref{sec:MCCQ-jets} was to introduce the
observer's trajectory, expand all fields in a Taylor series around it,
and quantize in the space of Taylor data histories. The motivation was
mainly aesthetic; by adding the observer's trajectory, it is possible to
write down a covariant expression (\ref{H2}) for the Hamiltonian, namely
as the operator which translates the fields relative to the observer.
However, it is in the presence of gauge symmetries that this
construction becomes indispensable.

Thus, we reformulate the classical theory in jet coordinates.
To the fields $c^a(x)$, $\fa(x)$, $\fsa(x)$ and $\za(x)$ we associates
$p$-jets $c^a_\cmm(t)$, $\fam(t)$, $\fsam(t)$ and $\zam(t)$, with
canonical momenta $b_a^\cmm(t)$, $\pam(t)$, $\psam(t)$ and
$\chi^{a\cmm}(t)$. We also introduce extra antifields $\wfam(t)$ etc.
to eliminate the $t$-dependence, but as in \cite{Lar05a}, they will not
be written down explicitly.

To be concrete, consider the case that the symmetry is the DGRO
algebra (\ref{DGRO}). To each symmetry, we assign ghosts as in the
following table:
\bes
\barr{|l|lllll|}
\hline
\hbox{} & \hbox{Gen} & \hbox{Smear}
& \hbox{Ghost} & \hbox{Momentum} & Q_{Long} \\
\hline
\hbox{Diffeomorphisms} & \L_\xi & \xmu(x)
& c_{diff}^\mu(x,t) & b^{diff}_\mu(x,t)
& Q^{diff}_{Long} \\
\hbox{Gauge} & \J_X & X^a(x)
& c_{gauge}^a(x,t) & b^{gauge}_a(x,t) & Q^{gauge}_{Long} \\
\hbox{Reparametrizations} & L_f & f(t)
&c_{rep}(t) & b^{rep}(t) & Q^{rep}_{Long} \\
\hline
\earr
\eens
The BRST operator is $Q_{BRST} = Q_{Long} + Q_{KT}$, where the
longitudinal operator is given by the prescription (\ref{smear}).
For brevity, we only write down the formulas for the fields $\fa(x,t)$
and the ghosts; the antifields do of course give rise to additional
terms.
\bes
Q^{diff}_{Long} &=& -\int \dNx \int dt\ \Big\{
(c_{diff}^\mu(x,t)\dmu\fa(x,t) \nl
&& +\dnu c_{diff}^\mu(x,t)T^{\al\nu}_{\bt\mu}\fb(x,t)) \pa(x,t) \nl
&&+ c_{diff}^\mu(x,t)\dmu c_{diff}^\nu(x,t)b^{diff}_\nu(x,t)
\Big\}, \nl
Q^{gauge}_{Long} &=& -\int \dNx\int dt\  \Big\{
c_{gauge}^a(x,t)J^{\al}_{\bt a}\fb(x,t) \pa(x,t) \nl
&&+\half f_{ab}{}^c c_{gauge}^a(x,t) c_{gauge}^b(x,t) b^{gauge}_c(x,t)
\Big\},
\label{QLongC}\\
Q^{rep}_{Long} &=& -\int \dNx \int dt\  \Big\{
(c_{rep}(t)\d_t\fa(x,t) \nl
&&+ \la(\dot c_{rep}(t)-ic_{rep}(t))\fa(x,t)) \pa(x,t) \Big\} \nl
&&- \int dt\ c_{rep}(t)\dot c_{rep}(t) b^{rep}(t).
\eens
These formulas assume that the field $\fa(x)$ transforms as a tensor
field. There is an additional term if the field is a connection, but
this terms does not lead to any complications.
After passage to jet space and normal ordering, we use the prescription
(\ref{smear}) to find the longitudinal derivative, i.e.
$\d_\mm \xmu \to c^\mu_{diff\cmm}$, $\d_\mm X^a \to c^a_{gauge\cmm}$,
and $f \to c_{rep}$:
\bes
Q^{diff}_{Long} &=& \int dt\ \Big\{
c^\mu_{diff,\bf0} p_\mu(t) - \sum_{|\nn|\leq|\mm|\leq p}
T^{\al\nn}_{\bt\mm}(c_{diff}(t)) \no{ \fbn(t)\pam(t) } \nl
&&-  \sum_{|\nn|\leq|\mm|\leq p} \no{ T^{\mu\nn}_{\nu\mm}(c_{diff}(t))
c^\nu_{diff,\nn}(t) b_\mu^{diff,\mm}(t) }
\Big\}, \nl
Q^{gauge}_{Long} &=& -\int dt\ \Big\{
\sum_{|\nn|\leq|\mm|\leq p}
J^{\al\nn}_{\bt\mm}(c_{gauge}(t)) \no{ \fbn(t)\pam(t) } \nl
&&- \half \sum_{|\nn|\leq|\mm|\leq p}
\no{ J^{a\nn}_{b\mm}(c_{gauge}(t))
c^b_{gauge,\nn}(t)b^{gauge,\mm}_a(t)}
\Big\},
\label{QLongJ}\\
Q^{rep}_{Long} &=& -\int dt\ \Big\{
\summ{p} c_{rep}(t)\no{\dotfam(t)\pam(t)} \nl
&& +\la\summ{p} (\dot c_{rep}(t)-ic_{rep}(t))\no{\fam(t)\pam(t)} \nl
&&+ \no{ c_{rep}(t)\dot c_{rep}(t) b^{rep}(t)}
\Big\}.
\eens
The matrices are given by (cf. (\ref{Tmn}))
\bes
T^\mm_\nn(c_{diff}(t)) &\equiv& (T^{\al\mm}_{\bt\nn}(c_{diff}(t)))
=\sum_{\mu\nu} {\nn\choose\mm} c^\mu_{diff,\nn-\mm+\nu}(t) T^\nu_\mu \nl
&& + \sum_\mu {\nn\choose\mm-\mu}c^\mu_{diff,\nn-\mm+\mu}(t)
 -\sum_\mu \dlt^{\mm-\mu}_\nn c^\mu_{diff,{\bf0}}(t), \nl
J^\mm_\nn(c_{gauge}(t)) &\equiv& (J^{\al\mm}_{\bt\nn}(c_{gauge}(t)))
= {\nn\choose\mm} c^a_{gauge,\nn-\mm}(t) J_a.
\ees
$T^{\mu\mm}_{\nu\nn}$ and $J^{a\mm}_{b\nn}$ denote the specializations
of $T^{\al\mm}_{\bt\nn}$ and $J^{\al\mm}_{\bt\nn}$ to the adjoint
representations;
$\sum_\mm T^{\mu\mm}_{\nu\nn}(c)c^\nu_\cmm = (c^\nu c^\mu_{,\nu})_\cnn$
and $\sum_\mm J^{a\mm}_{b\nn}(c)c^b_\cmm = (c^a c^b)_\cnn$

The condition for $Q_{Long}^2 = 0$, and thus $Q_{BRST}^2 = 0$, is
that the algebra generated by the normal-ordered gauge generators
is anomaly free.
However, even if this condition fails, which is the typical situation,
everything is not lost. The KT operator is still nilpotent, and
we can implement dyna\-mics as the KT cohomology in the extended phase space
without ghosts. The physical phase space now grows, because some gauge
degrees of freedom become physical upon quantization.

In the next two sections, we apply this formalism to some well-known
theories.

\section{The free Maxwell field}
\label{sec:Maxwell}

The Maxwell field $A_\mu(x)$ transforms as a vector field under the
Poincar\'e group and as a connection under the gauge algebra
$\map(N, \uu(1))$, whose smeared generators
are denoted by $\J_X = \int \dFx\  X(x) J(x)$:
\be
[\J_X, A_\mu(x)] = \dmu X(x).
\label{u1}
\ee
We use the Minkowski metric $\eta_{\mu\nu}$ and its inverse
$\eta^{\mu\nu}$ to freely raise and lower indices, e.g.
$F^{\mu\nu} = \eta^{\mu\rho}\eta^{\nu\si}F_{\rho\si}$.
The field strength $F_{\mu\nu}(x) = \dmu A_\nu(x) - \dnu A_\mu(x)$
transforms in the adjoint representation, i.e. trivially.
The action
\be
S = \quart \int \dFx\  F_{\mu\nu}(x)F^{\mu\nu}(x)
\ee
leads to the equations of motion
\be
\EE^\mu(x) \equiv -{\dlt S\/\dlt A_\mu(x)}
= \dnu F^{\mu\nu}(x) = 0.
\label{EMax}
\ee
The Maxwell equations are not all independent, because of the
constraints
\be
\dmu \EE^\mu(x) = \dmu\dnu F^{\mu\nu}(x) \equiv 0.
\label{cMax}
\ee
We are thus instructed to introduce
the following fields: the first-order antifield $\As^\mu(x)$
for the EL equation $\dnu F^{\mu\nu}(x) = 0$; the second-order antifield
$\zeta(x)$ for the identity $\dmu\dnu F^{\mu\nu}(x) \equiv 0$;
and the ghost $c(x)$ to identify fields related by a gauge transformation
of the form (\ref{u1}).

The BRST operator $s$ acts as
\bes
s c(x) &=& 0, \nl
s A_\mu(x) &=& \dmu c(x),
\nlb{sMax1}
s \As^\mu(x) &=& \dnu F^{\mu\nu}(x), \nl
s \zeta(x) &=& \dmu \As^\mu(x),
\eens
We check that $s^2 = 0$ and $s F_{\mu\nu} = s \dmu\As^\mu = 0$, so the
kernel of $s$ is spanned by $c$, the field strengths $F_{\mu\nu}$, and
$\dmu\As^\mu$. $\im s$ is generated by the ideals $\dmu c$,
$\dnu F^{\mu\nu}$, and $\dmu \As^\mu$. Hence $H_\cl^0(s)$
consists of the gauge-invariant parts of $A_\mu$ (i.e. $F_{\mu\nu}$)
which solve the Maxwell equations, as expected.

Introduce canonical momenta $E^\mu(x)$, $\Es_\mu(x)$, $\chi(x)$ and
$b(x)$, defined by the following non-zero brackets:
\bes
[E^\mu(x), A_\nu(x')] &=& \dlt^\mu_\nu \dlt(x-x'), \nl
\{\Es_\mu(x), \As^\nu(x')\} &=& \dlt_\mu^\nu \dlt(x-x'), \nle
{[}\chi(x), \zeta(x')] &=& \dlt(x-x'), \nl
\{b(x), c(x')\} &=&  \dlt(x-x').
\eens
It should be emphasized that $E^\mu = \dlt/\dlt A_\mu$ is the conjugate
of the gauge potential in history space, and not yet related to the
electric field $F^{\mu0}$. We could introduce the condition
$E^\mu \approx F^{\mu0}$ as a constraint in the history phase space,
turning the Maxwell equations into second class constraints.
By keeping dyna\-mics as a first-class constraint no such condition, which
would ruin covariance, is necessary.
The BRST operator can explicitly be written as
\be
Q_{BRST} = \int \dFx\  \Big\{ \dmu c(x)E^\mu(x)
+ \dnu F^{\mu\nu}(x)\Es_\mu(x) + \dmu \As^\mu(x)\chi(x) \Big\}.
\ee

The physical content of the theory is clearer in Fourier space.
The BRST operator
\be
Q_{BRST} &=& \int \dNk\ \Big\{ k_\mu c(k)E^\mu(-k)
+ (k^\mu k_\nu A^\nu(k) - k^\nu k_\nu A^\mu(k))\Es_\mu(-k) \nl
&&\qquad+ k_\mu \As^\mu(k)\chi(-k) \Big\},
\label{BRSTMaxw}
\ee
acts on the Fourier modes as
\bes
s c(k) &=& 0, \nl
s A_\mu(k) &=& k_\mu c(k), \nle
s \As^\mu(k) &=& k^\mu k_\nu A^\nu(k) - k^\nu k_\nu A^\mu(k), \nl
s \zeta(k) &=& k_\mu \As^\mu(k).
\eens
We distinguish between two cases:

1. $k^2 = \om^2 \neq 0$, say $k = (\om,0,0,0)$. Then
$sc = 0$, $s A_0 = \om c$, $s A_1 = s A_2 = s A_3 = 0$,
$s \As^0 = \om^2 A_0 - \om\om A_0 = 0$, $s\As^1 = \om^2 A_1$
$s\As^2 = \om^2 A_2$, $s\As^3 = \om^2 A_3$ and $s\zeta = \om \As^0$.
The kernel is thus spanned by $c$, $A_1$, $A_2$, $A_3$ and $\As^0$, and
the image is spanned by $c$, $A_1$, $A_2$, $A_3$ and $\As^0$. Since
$\ker s = \im s$ there is no cohomology.

2. $k^2 = 0$, say $k = (k_0,0,0,k_0)$. Then
$sc = 0$, $s A_0 = s A_3 = k_0 c$, $s A_1 = s A_2 = 0$,
$s \As^0 = s \As^3 = k^0 k_\nu A^\nu$, $s\As^1 = s\As^2 = 0$ and
$s\zeta = k_\mu \As^\mu$.
The kernel is thus spanned by $c$, $A_1$, $A_2$, $k^\mu A_\mu =
A_0 - A_3$, $\As^1$, $\As^2$, and  $k_\mu\As^\mu = \As^0 - \As^3$.
The image is spanned by $c$, $k^\mu A_\mu$ and $k_\mu\As^\mu$, which
factor out in cohomology. We are left with two physical polarizations
$A_1$ and $A_2$.

We here assumed that the momenta factor out in cohomology. As for the
scalar field in Section \ref{sec:Scalar1} and the harmonic oscillator
in \cite{Lar04}, this is not quite true. There is unwanted cohomology
because the Hessian is singular. However, this problem has nothing to
do with gauge invariance.

We now quantize in the history phase space before introducing dyna\-mics
by passing to the BRST cohomology.
We single out one direction $x^0$ as time, and take the Hamiltonian to
be the generator of rigid time translations,
\bes
H &=& -i\int \dFx\  \Big\{
  \d_0 A_\mu(x)E^\mu(x) + \d_0 \As^\mu(x)\Es_\mu(x) \nl
&&\qquad+ \d_0 \zeta(x)\chi(x) + \d_0 c(x)b(x) \Big\} \nle
&=& \int \dNk\ k_0 \Big\{
  A_\mu(k)E^\mu(-k) + \As^\mu(k)\Es_\mu(-k) \nl
&&\qquad+  \zeta(k)\chi(-k) + c(k)b(-k) \Big\}.
\eens
Note that at this stage we break Poincar\'e invariance,
since the Hamiltonian treats the $x^0$ coordinate differently from
the other $x^\mu$.
Quantize by introducing a Fock vacuum $\ket 0$ satisfying
\bes
&&A_\mu(k)\ket 0 = E^\mu(k)\ket 0 =
\As^\mu(k)\ket 0 = \Es_\mu(k)\ket 0 =
\nlb{LERA2}
&&\zeta(k)\ket 0 = \chi(k)\ket 0 =
c(k)\ket 0 = b(k)\ket 0 = 0,
\eens
for all $k$ such that $k_0 < 0$.

At this point we want to pass to BRST cohomology. There might be
problems with normal ordering, but in fact the BRST operator
(\ref{BRSTMaxw}) is already normal ordered. This is because
the generator of $\uu(1)$ gauge transformations
\be
\J_X = -\int \dFx\  X(x)\dmu E^\mu(x)
\label{u1gauge}
\ee
is itself already normal ordered. There are thus no anomalies, and
the BRST operator (\ref{BRSTMaxw}) remains nilpotent.
We define the BRST state cohomology as the space of physical states,
where a state is physical if it is BRST closed,
$Q_{BRST}\ket{phys} = 0$, and two physical states are equivalent if
they differ by a BRST exact state,
$\ket{phys} \sim \ket{phys'}$ if
$\quad \ket{phys} - \ket{phys'} = Q_{BRST}\ket{}$.

The rest proceeds as for the harmonic oscillator \cite{Lar04} or
the free scalar field in Section \ref{sec:Scalar1}.
After adding a small
perturbation to make the Hessian invertible, all momenta
vanish in cohomology, and only the transverse polarizations
$\eps^\mu A_\mu(k) = 0$ with $\eps^\mu k_\mu = 0$ and $\eps^0 = 0$
survive. A basis for the history Hilbert space consists of multi-quanta
states
\be
\eps^\mu_1A_\mu(k^{(1)})\ldots\eps^\mu_nA_\mu(k^{(n)})\ket{0}
\ee
where $k^{(j)}_\mu k^{(j)\mu} = 0$ and $k^{(j)}_0 > 0$.
The energy is given by $H = k_0^{(1)} + ...+ k_0^{(n)}$.
The gauge generators (\ref{u1gauge}) act in a well-defined manner,
in fact trivially, on the Hilbert space, because
$\eps^\mu_jk^{(j)}_\mu = 0$.

As in Section \ref{sec:MCCQ-jets}, we want to give a completely covariant
description of the Hamiltonian. Therefore we pass to jet data, e.g.
\be
A_\mu(x) = \summ{p} {1\/\mm!} A_{\mu\cmm}(t)(x-q(t))^\mm.
\label{ATaylor}
\ee
The equations of motion (\ref{EMax}) translate into
\be
\sum_\nu F^{\mu\nu}_{\cmm+\nu}(t) = 0,
\ee
and the constraint (\ref{cMax}) becomes
\be
\sum_\mu\EE^\mu_{\mu\cmm}(t)
= \sum_{\mu\nu} F^{\mu\nu}_{\cmm+\mu+\nu}(t)\equiv 0,
\ee
where the field strength is
\be
F_{\mu\nu\cmm}(t) = A_{\mu,\mm+\nu}(t) - A_{\nu,\mm+\mu}(t) = 0.
\ee
We introduce jets also for the antifields and for the ghost, denoted
by $A^\mu_{*,\mm}(t)$, $\zeta_{,\mm}(t)$, and
$c_{,\mm}(t) \equiv c^{gauge}_{,\mm}(t)$.
The BRST differential $s$ which implements all these conditions is
defined by
\bes
s c_{,\mm}(t) &=& 0, \nl
s A_{\mu,\mm}(t) &=& c_{,\mm+\mu}(t),
\nlb{sMax}
s A^\mu_{*,\mm}(t) &=& \sum_\nu F^{\mu\nu}_{,\mm+\nu}(t), \nl
s \zeta_{,\mm}(t) &=& \sum_\mu A^\mu_{*,\mm+\mu}(t).
\eens

Moreover, we demand that the Taylor series does not depend on the
parameter $t$, which gives rise to conditions of the type
\be
D_t A_{\mu,\mm}(t) \equiv \dot A_{\mu,\mm}(t)
- \sum_\nu \dot q^\nu(t)A_{\mu,\mm+\nu}(t) &=& 0.
\ee
As in (\ref{dltsi}), we need to double the number of antifields
and introduce an additional differential $\sigma$ to remove these
conditions in cohomology. Thus we introduce antifields
$\bar c_{,\mm}(t)$, $\bar A_{\mu,\mm}(t)$, $\bar A^\mu_{*,\mm}(t)$,
$\bar \zeta_{,\mm}(t)$ and set
\bes
\sigma \bar c_{,\mm}(t) &=& \dot c_{,\mm}(t), \nl
\sigma \bar A_{\mu,\mm}(t) &=& \dot A_{\mu,\mm}(t), \nl
\sigma \bar A^\mu_{*,\mm}(t) &=& \dot A^\mu_{*,\mm}(t), \\
\sigma \bar \zeta_{,\mm}(t) &=& \dot \zeta_{,\mm}(t) \nl
\sigma c_{,\mm}(t) &=& \sigma A_{\mu,\mm}(t) = \sigma A^\mu_{*,\mm}(t)
= \sigma \zeta_{,\mm}(t) = 0.
\eens
Clearly, $\sigma^2 = 0$.
We also extend the definition of the BRST differential $s$ to the
barred antifields:
\bes
s \bar c_{,\mm}(t) &=& 0, \nl
s \bar A_{\mu,\mm}(t) &=& -\bar c_{,\mm+\mu}(t), \nle
s \bar A^\mu_{*,\mm}(t) &=&
-\sum_\nu (\bar A_{\nu,\mm+\mu}(t) - \bar A_{\mu,\mm+\nu}(t)), \nl
s \bar \zeta_{,\mm}(t) &=& -\sum_\mu \bar A^\mu_{*,\mm+\mu}(t).
\eens
That $s^2 = 0$ follows in the same way as for (\ref{sMax}).
Moreover, we verify that $s\sigma = -\sigma s$, and hence $s+\sigma$
is nilpotent.

The classical cohomology group $H_\cl^0(s+\sigma)$ consists of linear
combinations of jets
\be
A_{\mu,\mm}(t) = \eps_\mu(t) \e^{ik\cdot q(t)} (ik)^\mm
\ee
where $k^2=0$ and the polarization vector $\eps_\mu(t)$ is perpendicular
both to the photon momentum and the observer's trajectory:
\be
\eps_\mu(t) k^\mu = \eps_\mu(t) \dot q^\mu(t) = 0.
\ee
The latter is evidently equivalent to the non-covariant condition
$\eps^0 = 0$. Moreover, $k\cdot q = k_\mu q^\mu$.
The Taylor series (\ref{ATaylor}) can be summed in the same way as for
the scalar field (\ref{phixt}).

We now quantize the theory before imposing dyna\-mics. To this end,
we introduce the canonical momenta for all jets and antijets, and
$p_\mu(t)$ and $\psmu(t)$ for the observer's trajectory and its
antifield. The defining relations are
\bes
[E^{\mu,\mm}(t), A_{\nu,\nn}(t')]
&=& \dlt^\mu_\nu \dlt^\mm_\nn \dlt(t-t'), \nl
\{E^{*\cmm}_\mu(t), A^\nu_{*\cnn}(t')\}
&=& \dlt_\mu^\nu \dlt^\mm_\nn \dlt(t-t'), \nl
{[}\chi^\cmm(t), \zeta_\cnn(t')] &=& \dlt^\mm_\nn \dlt(t-t'), \\
\{b^\cmm(t), c_\cnn(t')\} &=&  \dlt^\mm_\nn \dlt(t-t'), \nl
{[}p_\nu(t),  q^\mu(t')] &=& \dlt^\mu_\nu \dlt(t-t').
\eens
Since the jets also depend on the parameter $t$, we can define
their Fourier components as in (\ref{Fourier}).
The Fock vacuum (\ref{LERA2}) is replaced by a new vacuum, also denoted
by $\ket 0$, which is defined to be annihilated by
the negative frequency modes.
The quantum Hamiltonian is still defined by (\ref{Hq}),
where double dots indicate normal ordering with respect to frequency,
ensuring that $H\ket 0 = 0$.

It remains to check that the algebra of $\uu(1)$ gauge transformations
acts in a well-defined manner before we can pass to the BRST cohomology.
Since a gauge potential transforms as
\be
[\J_X, A_{\mu,\mm}(t)] = \d_{\mm+\mu}X(q(t))
\ee
we have
\be
\J_X = \summ{p} \sum_\mu \int dt\ \d_{\mm+\mu}X(q(t))E^{\mu\cmm}(t).
\label{JXMax}
\ee
There are no contributions from the antifields, since $\As^\mu$,
$\zeta$ and $c$ all transform trivially under $\map(N,\uu(1))$.
The prescription (\ref{smear}) gives
\be
Q_{Long} = \summ{p} \sum_\mu \int dt\ c_{\cmm+\mu}(t)E^{\mu\cmm}(t).
\label{QLMax}
\ee
The expressions (\ref{JXMax}) and (\ref{QLMax}) are evidently normal
ordered as they stand, and  consequently there are no gauge anomalies.

The rest proceeds as for the scalar field.

\section{Gravity}
\label{sec:Gravity}

Finally we are ready to apply the MCCQ formalism to general relativity.
For simplicity we consider only pure gravity.
The only field is the symmetric metric $g_{\mu\nu}(x)$. The
inverse $g^{\mu\nu}$, the determinant $g = \det (g_{\mu\nu})$,
the Levi-Civit\`a connection
$\Gamma^\mu_{\nu\rho}$, Riemann's curvature tensor
$ R^\rho{}_{\si\mu\nu}$, the Ricci tensor $R_{\mu\nu}$, the
scalar curvature $R = g^{\mu\nu}R_{\mu\nu}$ and the Einstein
tensor $G^{\mu\nu} = R^{\mu\nu} - \half g^{\mu\nu} R$
are defined as usual.
The covariant derivative is
\be
\nabla_\mu = \dmu + \Gamma^\rho_{\nu\mu} T^\nu_\rho,
\ee
where $T^\mu_\nu$ are finite-dimensional matrices satisfying $gl(N)$
(\ref{glN}).

The Einstein action
\be
S_E = {1\/16\pi} \int \dFx\  \sqrt{g(x)} R(x).
\ee
leads to Einstein's equation of motion
\be
G^{\mu\nu}(x) = 0,
\label{Einstein}
\ee
which is subject to the identity
\be
\nabla_\nu G^{\mu\nu}(x) \equiv 0
\label{Econt}
\ee
We introduce a fermionic antifield $g^{\mu\nu}_*(x)$ for (\ref{Einstein}),
a bosonic second-order antifield $\zeta^\mu(x)$ for (\ref{Econt}), and
a ghost $c^\mu_{diff}(x)$ to eliminate diffeomorphisms. The total field
content in the extended history phase space is thus
\bes
\barr{|l|lll|}
\hline
\hbox{afn} & \hbox{Field} & \hbox{Momentum} & \hbox{Parity}\\
\hline
-1 & c^\mu_{diff}(x) & b_\mu^{diff}(x) & F\\
0 & g_{\mu\nu}(x) & \pi^{\mu\nu}(x) &B \\
1 & g^{\mu\nu}_*(x) & \pi^*_{\mu\nu} &F\\
2 & \zeta^\mu(x) & \chi_\mu(x) &B\\
\hline
\earr
\eens
The KT differential $\dlt$ is defined by
\bes
\dlt c^\mu_{diff}(x) &=& 0, \nl
\dlt g_{\mu\nu}(x) &=& 0, \nle
\dlt g^{\mu\nu}_*(x) &=&  G^{\mu\nu}(x) \nl
\dlt \zeta_\mu(x) &=& \nabla_\nu G^{\mu\nu}(x),
\eens
i.e. the KT operator is
\be
Q_{KT} = \int \dFx\ \ (G^{\mu\nu}(x)\pi^*_{\mu\nu}(x)
+ \nabla_\nu G^{\mu\nu}(x)\chi_\mu(x)).
\ee
The longitudinal operator was written down in (\ref{QLongC}),
$Q_{Long} = Q_{Long}^{diff}$, and the BRST operator is the sum of the
KT and the longitudinal operators, as usual. Of the originally ten
degrees of freedom $g_{\mu\nu}(x)$, the antifield $g^{\mu\nu}_*(x)$
eliminates four and the ghost $c^\mu_{diff}(x)$ another four,
leaving two graviton polarizations in the BRST cohomology.

We now quantize as usual by passing to jet space, introducing a Fock
vacuum that is annihilated by the negative frequency modes of all fields
and antifields, and normal ordering.
The fields are symmetric tensor fields, i.e. they correspond to the
symmetric $gl(N)$ modules $S_\ell$ and $S^\ell$ in (\ref{kSl}).
The values of the parameters in (\ref{numdef}) are
\bes
\barr{|l|c|cccc|c|}
\hline
\hbox{Field} & \rep & u & v & w & x & p\\
\hline
c^\mu_{diff}(x) & S^1 & 1 & 0 & -1 & N & p+1 \\
g_{\mu\nu}(x) & S_2 & -(N+2) & -1 & -(N+1) & -N(N+1)/2 & p  \\
g^{\mu\nu}_*(x) & S^2 & (N+2) & 1 & -(N+1) & N(N+1)/2 & p-2 \\
\zeta^\mu(x) & S^1 & -1 & 0 & 1 & -N & p-3 \\
\hline
\earr
\eens
The parameters were written down in arbitrary dimension $N$ for
generality, although we are primarily interested in the physical case
$N=4$.
The last column is the truncation order for the corresponding jets.

The diffeomorphism anomalies are now read off from (\ref{cs}); for
definiteness, we only consider $c_1$. Depending
on whether we exclude the ghost $c^\mu_{diff}(x)$ or not, the abelian
charge $c_1 = c_1^{tot}$ becomes
$c^{tot}_1 = c^{field}_1 \equiv 1 + c^g_1 + c^{g*}_1 + c^\zeta_1$
or $c^{tot}_1 = c^{field}_1 + c^{ghost}_1$, where
\bes
c^{ghost}_1 &=& - {N+p+1\choose N} - N {N+p+2\choose N+2}, \nl
c^g_1 &=& (N+2){N+p\choose N} + {N(N+1)\/2}{N+p+1\choose N+2}, \nl
c^{g*}_1 &=& - (N+2) {N+p-2\choose N}
- {N(N+1)\/2} {N+p-1\choose N+2}, \nl
c^\zeta_1 &=&  {N+p-3\choose N} +N {N+p-2\choose N+2}.
\ee
It is clear that $c^{tot}_1$ does not vanish for generic $p$, with
or without the ghost contribution (the leading term proportional to
$p^{N+2}/(N+2)!$ does however vanish in the $p\to\infty$ limit). The
longitudinal operator (\ref{QLongJ}) thus acquires an anomaly, and we
can only implement the KT cohomology. Hence the ghost plays no role and
should be discarded.

The quantum KT operator becomes
\bes
Q_{KT} &=& \int dt\ \Big\{
\summ{p-2} \no{ G^{\mu\nu}_\cmm(t)\pi^{*\cmm}_{\mu\nu}(t) } \nle
&&+ \summ{p-3} \no{ (\nabla_\nu G^{\mu\nu})_\cmm(t)\chi^\cmm_\mu(t) }
\Big\},
\eens
where $G^{\mu\nu}_\cmm(t)$ and $(\nabla_\nu G^{\mu\nu})_\cmm(t)$
are the corresponding jets. Note that the sums run up to
$|\mm| = p-2$ and $p-3$, respectively, because Einstein's equation
is second order and the identity (\ref{Econt}) is third order.

One difference compared to Minkowski space is that the geodesic equation
depends on a dynamical field. In order to make the geodesic
operator $\GG_\mu(t)$ transform tensorially under repara\-metrizations as
well, we need to add an extra term. We can construct the following
quantites from the metric and the observer's trajectory:
\begin{enumerate}
\item
The Levi-Civit\`a connection
$\Gamma^\nu_{\si\tau}(x,t) = \half g^{\nu\rho}(x,t)(\d_\si
g_{\rho\tau}(x,t) +
\break \d_\tau g_{\si\rho}(x,t) - \d_\rho g_{\si\tau}(x,t))$.
\item
The einbein $e(t) =
\sqrt{g_{\mu\nu,\bf0}(t) \dot q^\mu(t)\dot q^\nu(t)}$.
\item
The repara\-metrization connection $\Gamma(t) = -e^{-1}(t)\dot e(t)$.
\end{enumerate}
The geodesic operator reads \cite{Lar02}
\be
\GG_\mu(t) =
e^{-1}(t) g_{\mu\nu,{\bf0}}(t)
( \ddot q^\nu(t) + \Gamma(t) \dot q^\nu(t)
+ \Gamma^\nu_{\si\tau,{\bf0}}(t)\dot q^\si(t)\dot q^\tau(t) ),
\label{ggeo}
\ee
where $g_{\mu\nu,{\bf0}}(t)$ and $\Gamma^\nu_{\si\tau,{\bf0}}(t)$
are the zero-jets
corresponding to the metric and Levi-Civit\`a connection, respectively.
It is straightforward to check that (\ref{ggeo}) transforms nicely
under the full DGRO algebra,
\bes
{[}\L_\xi, \GG_\nu(t)] &=& -\dnu\xmu(q(t))\GG_\mu(t), \nle
{[}L_f, \GG_\nu(t)] &=& -f(t)\dot\GG_\nu(t) - \dot f(t)\GG_\nu(t).
\eens
The contribution to the KT operator is thus
$Q_{KT} = \int dt\ \GG_\mu(t)\psmu(t)$,
which eliminates the observer's trajectory $q^\mu(t)$ in cohomology.

\section{Finiteness conditions}
\label{sec:Finiteness}

In the previous section we applied the MCCQ formalism to gravity, and
found a well-defined but anomalous action of the DGRO algebra. However,
the passage to the space of $p$-jets amounts to a regularization. The
regularization is unique in that it preserves the full constraint
algebra, but it must nevertheless be removed in the end. In order to
reconstruct the original field by means of the Taylor series
(\ref{Taylor}), we must take the limit $p\to\infty$. A necessary
condition for taking this limit is that the abelian charges have a
finite limit.

Taken at face value, the prospects for succeeding
appear bleak. When $p$ is large, ${m+p\choose n} \approx p^n/n!$, so
the abelian charges (\ref{cs}) diverge; the worst case is
$c_1 \approx c_2 \approx p^{N+2}/(N+2)!$, which diverges in all dimensions
$N > -2$. In \cite{Lar01} a way out of this problem was devised: consider
a more general realization by taking the direct sum of operators
corresponding to different values of the jet order $p$. Take
the sum of $r+1$ terms like those in (\ref{Fock}), with $p$ replaced by
$p$, $p-1$, ..., $p-r$, respectively, and with $\rep$ and $M$ replaced by
$\repi$ and $\Mi$ in the $p-i$ term.

Such a sum of contributions arises naturally from the KT complex, because
the antifields are only defined up to an order smaller than $p$ (e.g.
$p-\ord_\al$ or $p-\ordg_a$).
Denote the numbers $u,v,w,x,y$ in the modules $\repi$ and $\Mi$,
defined as in (\ref{numdef}), by $u_i,v_i,w_i,x_i,y_i$,
respectively. Of course, there is
only one contribution from the observer's trajectory.
Then it was shown in \cite{Lar01}, Theorem 3, that
\bes
c_1 = -U\Npr, &\qquad&
c_2 = -V\Npr, \nl
c_3 = W\Npr, &\qquad&
c_4 = -X\Npr,
\label{finc} \\
c_5 = Y\Npr,
\eens
where $u_0=U$, $v_0=V$, $w_0=W$, $x_0=X$ and $y_0=Y$, provided that the
following conditions hold:
\bes
&&u_i + \ritwo X = \mri U, \nl
&&v_i - 2\rione W - \ritwo X = \mri V, \nl
&&w_i - \rione X = \mri W,
\label{conds}\\
&&x_i = \mri X, \nl
&&y_i = \mri Y.
\eens
The contributions from the observer's trajectory have also been eliminated
by antifields coming from the geodesic equation; this is
not important in the sequel because these contributions were finite
anyway.

Let us now consider the solutions to (\ref{conds}) for the numbers
$x_i$, which can be interpreted as the number of fields and anti-fields.
First assume
that the field $\fam(t)$ is fermionic with $x_F$ components, which gives
$x_0=x_F$. We may assume, by the spin-statistics theorem, that the
EL equations are first order, so the bosonic antifields $\fsam(t)$
contribute $-x_F$ to $x_1$. The barred antifields $\wfam(t)$ are also
defined up to order $p-1$, and so give $x_1=-x_F$, and the barred
second-order antifields $\wfsam(t)$ give $x_2=x_F$. Further assume that
the fermionic EL equations have $x_S$ gauge symmetries, i.e. the
second-order antifields $\zam(t)$ give $x_2=x_S$. In established
theories, $x_S=0$, but we will need a non-zero value for $x_S$.
Finally, the corresponding barred antifields give $x_3 = -x_S$.

For bosons the situation is analogous, with two exceptions: all signs
are reversed, and the EL equations are assumed to be second order.
Hence $\fsam(t)$ yields $x_2=x_B$ and the gauge antifields $\zam(t)$
give $x_3=-x_G$. Accordingly, the barred antifields are one order higher.

The situation is summarized in the following tables, where the upper half
is valid if the original field is fermionic and the lower half if it is
bosonic:
\bes
\barr{|c|c|c|l|}
\hline
\afn & \hbox{Jet} & \hbox{Order} &x \\
\hline
0 & \fam(t) & p & x_F \\
1 & \wfam(t) & p-1 & -x_F \\
1 & \fsam(t) & p-1 & -x_F \\
2 & \wfsam(t) & p-2 & x_F \\
2 & \zam(t) & p-2 & x_S \\
3 & \bar\zam(t) & p-3 & -x_S \\
\hline
\hline
0 & \fam(t) & p & -x_B \\
1 & \wfam(t) & p-1 & x_B \\
1 & \fsam(t) & p-2 & x_B \\
2 & \wfsam(t) & p-3 & -x_B \\
2 & \zam(t) & p-3 & -x_G \\
3 & \bar\zam(t) & p-4 & x_G \\
\hline
\earr
\label{tab}
\ees
If we add all contributions of the same order, we see that fourth
relation in (\ref{conds}) can only be satisfied provided that
\bes
p: &\quad& x_F - x_B = X \nl
p-1: && -2x_F+x_B = -rX, \nl
p-2: && x_B + x_F + x_S = {r\choose2}X, \nl
p-3: && -x_B-x_S-x_G = -{r\choose3}X,
\label{rcond}\\
p-4: && x_G = {r\choose4}X, \nl
p-5: && 0 = -{r\choose5}X, ...
\eens
The last equation holds only if $r\leq4$ (or trivially if $X=0$).
On the other hand, if we demand that there is at least one bosonic
gauge condition, the $p-4$ equation yields $r\geq4$. Such a demand is
natural, because both the Maxwell/Yang-Mills and the Einstein equations
have this property. Therefore, we are unambigiously guided to consider
$r=4$ (and thus $N=4$). The specialization of (\ref{rcond}) to four
dimensions reads
\bes
p: &\quad& x_F - x_B = X \nl
p-1: && -2x_F+x_B = -4X, \nl
p-2: && x_B + x_F + x_S = 6X, \\
p-3: && -x_B-x_S-x_G = -4X, \nl
p-4: && x_G = X.
\eens
Clearly, the unique solution to these equations is
\be
x_F = 3X, \qquad x_B = 2X, \qquad x_S = X, \qquad x_G = X.
\label{xsol}
\ee
The solutions to the remaining equations in (\ref{conds}) are found by
analogous reasoning. The result is
\bes
\barr{llllll}
u_B = 2U  &\qquad& v_B = 2V+2W \\
u_F = 3U  && v_F = 3V+2W \\
u_S = U-X && v_S = V+2W+X \\
u_G = U-X && v_G = V+2W+X \\
\\
w_B = 2W+X && y_B = 2Y \\
w_F = 3W+X && y_F = 3Y \\
w_S = W+X  && y_S = Y \\
w_G = W+X  && y_G = Y \\
\\
\earr
\ees
This result expresses the twenty parameters $x_B-w_G$
in terms of the five parameters $X$, $Y$, $U$, $V$, $W$.
For this particular choice of parameters, the abelian charges in
(\ref{finc}) are given by
\be
c_1 = -U, \qquad c_2 = -V, \qquad c_3 = W, \qquad c_4 = -X,
\qquad c_5 = Y,
\ee
independent of $p$. Hence there is no manifest obstruction to the
limit $p\to\infty$.

The prediction that spacetime has $N=4$ dimensions is of course very
nice. Unfortunately, at closer scrutiny the situation appears less
appealing. In particular, the need for fermionic gauge symmetries ($x_S
\neq 0$) is apparently in disagreement with observation. It was also
found in \cite{Lar05a} that the gauge anomaly $c_5$ does not have a
finite $p\to\infty$ limit for reasonable choices of field content.

Hence it is presently unclear how to remove the regulator and take the
field limit, and this is of course a major unsolved
problem. Nevertheless, it should be emphasized that already the
regularized theories carry representations of the {\em full} gauge and
diffeomorphism algebras.

\section{Conceptual issues}
\label{sec:Concept}

One of the most important tasks of any putative quantum theory of
gravity is to shed light on the various conceptual difficulties which
arise when the principles of quantum mechanics are combined with
general covariance \cite{Car01,Sav04}. These issues include:
\begin{enumerate}
\item
In conventional canonical quantization, the canonical commutation
relations are defined on a ``spacelike'' surface. However, a surface is
spacelike w.r.t. some particular spacetime metric $g_{\mu\nu}$, which is
itself a quantum operator.
\item
Microcausality requires that the field variables
defined in spacelike separated regions commute. Again, it is unclear
what this means when the notion of spacelikeness is dynamical.
\item
Different choices of foliation lead to a priori different quantum
theories, and it by no means clear that these are unitarily equivalent.
\item
The problem of time: The Hamiltonian of general relativity is a first
class constraint, hence it vanishes on the reduced phase space. This
means that there is no notion of time evolution among
diffeomorphism-invariant degrees of freedom.
\item
The notion of time as a causal order is lost. This is not really a
problem in the classical theory, where one can solve the equations of
motion first, but in quantum theory causality is needed from the outset.
\item
QFT rests on two pillars: quantum mechanics and locality. However,
locality is at odds with diffeomorphism invariance underlying gravity;
``there are no local observables in quantum gravity''.
\end{enumerate}

Let us see how MCCQ addresses these conceptual issues.
\begin{enumerate}
\item
The canonical commutation relations are defined throughout the history
phase space $\PP$, and hence not restricted to variables living on a
spacelike surface. Dynamics is implemented as a first class constraint
in $\PP$. Only if we solve this constraint prior to quantization need we
restrict quantization to a spacelike surface.
\item
By passing to $p$-jet space, we eliminate the notion of spacelikeness
altogher. The $p$-jets live on the observer's trajectory, and the
observer moves along a timelike curve. It
might seem strange to dismiss the notion of spacelike separation, but
distant events can never be directly
observed, and a physical theory only needs to describe directly
observable events. What can be observed are indirect effects of distant
events. E.g., a terrestial detector does not directly observe the sun,
but only photons emanating from the sun eight light-minutes ago. The
detector signals are of course compatible with the existence of the sun,
but a physical theory only needs to deal with directly observed events,
i.e. the absorbtion of photons in the detector.
\item
In MCCQ there is no foliation, but rather an explicit observer, or
detector. The theory is
unique since the observer's trajectory is a quantum object; we do not
deal with a family of theories parametrized by the choice of observer,
but instead the observer's trajectory is represented on the Hilbert
space in the same way as the quantum fields.
\item
By introducing an explicit observer, we can define a genuine energy
operator (\ref{Hq}) which translates the fields relative to the observer,
or vice versa. In contrast, there is also a Hamiltonian constraint, which
translates both the observer and the fields the same amount. This
constraint is killed in KT cohomology and is thus identically zero on
physical observables.
\item
The $p$-jets live on the observer's trajectory $q(t)$ and are
thus causally related; causal order is defined by the parameter $t$.
The relation between this order and the fields is encoded in the
geodesic equation (\ref{ggeo}).
\item
As we saw in Section \ref{sec:Locality}, locality is compatible with
infinite-dimensional spacetime symmetries, but only in the presence of
an anomaly. This is the key lesson from CFT.
\end{enumerate}
It is gratifying that the MCCQ formalism yields natural explanations of
many of the conceptual problems that plague quantum gravity.

\section{Conclusion}
\label{sec:Conclusion}

The key insight underlying the present work is that the process of
observation must be localized in spacetime in order to be compatible with
the philosophy of QFT. The innocent-looking introduction of the
observer's trajectory leads to dramatic consequences, because new gauge
and diffeomorphism anomalies arise. On the mathematical side, this
construction leads to well-defined realizations of the constraint algebra
generators as operators on a linear space, as least for
the regularized theory.

We have also developed a manifestly covariant canonical quantization
method, based on the form of the DGRO algebra modules. This formalism
is convenient due to its relation to representation theory, but it is
presumably possible to repeat the analysis in any sensible quantization
scheme, at the cost of additional work. In contrast, the introduction
of the observer's trajectory is absolutely crucial, because the new
anomalies can not be formulated without it. Anomalies matter!

Four critical problems remain to be solved. As was discussed in Section
\ref{sec:Finiteness}, the original fields must be reconstructed from the
$p$-jets, i.e. we must take the limit $p\to\infty$. This limit is
problematic because the abelian charges diverge. Second, the issue of
unitarity needs to be understood. So far we only noted that an extension
is necessary for unitarity by restriction to Virasoro subalgebras, and
then we proceeded to construct anomalous representations. The main
problem is to find an invariant inner product. Third, perturbation
theory and renormalization must be transcribed to ths formalism, to make
contact with numerical predictions of ordinary QFT. Finally, we know
from CFT that reducibility conditions analogous to Kac' formula
\cite{FMS96} are needed in physically interesting situations.
Unfortunately, none of these problems appears to be easy.

\end{document}